\documentclass{article}

\usepackage{PRIMEarxiv}

\usepackage[utf8]{inputenc} 
\usepackage[T1]{fontenc}    
\usepackage{hyperref}       
\usepackage{url}            
\usepackage{booktabs}       
\usepackage{amsfonts}       
\usepackage{nicefrac}       
\usepackage{microtype}      
\usepackage{lipsum}
\usepackage{fancyhdr}       
\usepackage{graphicx}       
\usepackage{amsmath}
\usepackage{multirow}
\usepackage{multicol}
\usepackage{array}
\graphicspath{{media/}}     

\title{Unraveling Radiomics Complexity: Strategies for Optimal Simplicity in Predictive Modeling}

\author{
  \textbf{Mahdi Ait Lhaj Loutfi$^1$, Teodora Boblea Podasca$^2$, Alex Zwanenburg$^{3,4,5,6}$, Taman Upadhaya$^7$, Jorge Barrios Ginart$^8$, David R Raleigh$^{9}$, William C. Chen$^8$, Dante P.I. Capaldi$^8$, Hong Zheng$^{10}$, Olivier Gevaert$^{11}$, Jing Wu$^{12}$, Alvin C. Silva$^{13}$, Paul J. Zhang$^{14}$, Harrison X. Bai$^{15}$, Jan Seuntjens$^{16}$, Steffen Löck$^3$, Patrick O. Richard$^{17}$, Olivier Morin$^8$, Caroline Reinhold$^{18}$, Martin Lepage$^{19,20}$, Martin Vallières$^{1,21,*}$}\\
  \vskip 0.2in \center{}   \vskip 0.2in
  $^1$Department of Computer Science, Université de Sherbrooke, Sherbrooke, QC, Canada
  \\$^2$Department of Surgery, Service of Urology, Université de Sherbrooke, Sherbrooke, QC, Canada
  \\$^3$OncoRay – National Center for Radiation Research in Oncology, Faculty of Medicine and University Hospital Carl Gustav Carus, TUD Dresden University of Technology, Helmholtz-Zentrum Dresden-Rossendorf, Dresden, Germany
  \\$^4$National Center for Tumor Diseases Dresden (NCT/UCC), Germany: German Cancer Research Center (DKFZ), Heidelberg, Germany
  \\$^5$Faculty of Medicine and University Hospital Carl Gustav Carus, TUD Dresden University of Technology, Dresden, Germany
  \\$^6$Helmholtz-Zentrum Dresden-Rossendorf (HZDR), Dresden, Germany
  \\$^7$Department of Radiation Oncology, Cedars-Sinai Medical Center, Los Angeles, CA, USA
  \\$^8$Department of Radiation Oncology, University of California San Francisco, San Francisco, CA, USA
  \\$^9$Departments of Radiation Oncology, Neurological Surgery, and Pathology, University of California San Francisco, San Francisco, CA, USA
  \\$^{10}$Center for Biomedical Informatics Research, School of Medicine, Stanford University, CA 94305, USA
  \\$^{11}$Department of Medicine, and Department of Biomedical Data Science, Stanford University, Stanford, CA 94305, USA
  \\$^{12}$Department of Radiology, The Second Xiangya Hospital of Central South University, Changsha, 410011, Hunan, China
  \\$^{13}$Department of Radiology, Mayo Clinic Arizona, 13400 E Shea Blvd., Scottsdale, AZ 85259, USA
  \\$^{14}$Department of Pathology and Clinical Medicine, Hospital of the University of Pennsylvania, Philadelphia, PA, USA
  \\$^{15}$Department of Radiology and Radiological Sciences, Johns Hopkins, Baltimore, MD, USA
  \\$^{16}$Princess Margaret Cancer Centre, University Health Network \& Departments of Radiation Oncology \& Medical Biophysics, University of Toronto, Toronto, ON, Canada.
  \\$^{17}$Division of Urology, Centre Hospitalier Universitaire de Sherbrooke; Université de Sherbrooke Cancer Research Institute, Sherbrooke, QC, Canada
  \\$^{18}$Department of Radiology, McGill University Health Center, Director and Co-founder, Augmented Intelligence Precision Health Laboratory (AIPHL) of the Research Institute of the McGill University Health Center, Montreal, QC, Canada
  \\$^{19}$Département de médecine nucléaire et radiobiologie, Université de Sherbrooke, Sherbrooke, QC, Canada
  \\$^{20}$Centre d'imagerie moléculaire de Sherbrooke, Université de Sherbrooke, Sherbrooke, QC, Canada
  \\$^{21}$Centre de recherche du Centre hospitalier universitaire de Sherbrooke, Université de Sherbrooke, Sherbrooke, QC, Canada.
  \\$^{*}$Corresponding author: \texttt{martin.vallieres@usherbrooke.ca}
}

\begin{document}
\maketitle
\vskip 1.2in
\begin{abstract}

\textbf{Background}: The high dimensionality of radiomic feature sets, the variability in radiomic feature types and potentially high computational requirements all underscore the need for an effective method to identify the smallest set of predictive features for a given clinical problem.

\textbf{Purpose}: To establish a methodology and provide tools for identifying and explaining the smallest set of predictive features radiomic features.

\textbf{Materials and Methods}: Radiomic features (a total of 89,714) were extracted from five distinct datasets with different cancer types: low-grade glioma, meningioma, non-small cell lung cancer (NSCLC), and two renal cell carcinoma cohorts (n=2104). These features were categorized into complexity levels, defined by the number of computational steps required for their computation, encompassing morphological, intensity, texture, linear filters-based, and nonlinear filter-based features. For every dataset, models were trained on each complexity level specifically to classify clinical outcomes, and their performance was evaluated using the area under the curve (AUC). The most informative features were identified and their importance was explained. The optimal complexity level and associated most informative features were identified using systematic statistical significance analyses and a false discovery avoidance procedure, respectively. Their predictive importance was explained using a novel tree-based method.

\textbf{Results}: MEDimage, a new open-source tool, was designed and implemented to streamline radiomic studies through both code-based and graphical-based approaches, and was applied using our proposed methodology to analyze the datasets. Morphological features were found to be optimal in two cases: for MRI-based meningioma (AUC: 0.65; sensitivity: 64\%; specificity: 62\%; 95\% CI: 0.59, 0.72) and MRI-based low-grade glioma (AUC: 0.68; sensitivity: 68\%; specificity: 69\%; 95\% CI: 0.60, 0.75). Additionally, intensity features were optimal in two instances: for contrast-enhanced CT (CECT)-based renal cell carcinoma (AUC: 0.82; sensitivity: 77\%; specificity: 78\%; 95\% CI: 0.76, 0.88) and CT-based NSCLC (AUC: 0.76; sensitivity: 73\%; specificity: 71\%; 95\% CI: 0.71, 0.80). Texture features were identified as optimal for MRI-based renal cell carcinoma (AUC: 0.72; sensitivity: 71\%; specificity: 65\%; 95\% CI: 0.68, 0.77). Notably, in CECT-based renal cell carcinoma, the tuning of the Hounsfield unit range, which directly affects intensity-based features, led to improved results (AUC: 0.86).

\textbf{Conclusion}: Our proposed methodology and software can estimate the optimal radiomics complexity level for specific medical outcomes, potentially simplifying the use of radiomics in predictive modeling across various contexts.
\end{abstract}


\section{Introduction}
Medical imaging is a cornerstone of personalized medicine by providing a non-invasive window into the unique phenotypic characteristics of volumes of interest. Radiomics is defined as the high-throughput extraction of quantitative features from images to enable characterization of tissues \cite{images_more_data}. Such features extend our ability to discern subtle nuances in medical imaging characteristics, providing data for predictive modeling approaches to enhance personalized treatment \cite{radiomics_bridge}.

Radiomic analysis encompasses various feature categories, each potentially related to different aspects of a tissue phenotype. These categories include morphological features, which describe the shape and size of analyzed regions; intensity features, which capture characteristics related to pixel intensity distributions; and texture features, which quantify spatial intensity patterns. These features can be extracted from the original image intensities (e.g., Hounsfield Units in x-ray CT or arbitrary intensities in MRI), but also from image intensities previously modified by filters that highlight various structures and patterns \cite{ibsi_2}.

Indeed, different image pre-processing steps such as interpolation and intensity range definition may be performed prior to feature extraction. Some radiomic features, such as the mean of gray level intensities, can be simple quantities to calculate. Others, such as texture features are more complex constructs and require more computational steps. Many features also rely on adjustable parameters, such as the intensity discretisation scheme, prior to texture calculations. Recently, the Image Biomarker Standardization Initiative (IBSI) defined a standardized workflow for radiomic feature extraction, with and without filters \cite{ibsi_1, ibsi_2}. The IBSI also established reference values for 173 features and eight linear filters (LF), which can be used to calibrate radiomic software. Nonetheless, it is up to research teams to define the set of radiomic features and associated extraction parameters most relevant to address a given clinical question. With multiple options for image pre-processing steps and adjustable parameters, a given radiomic feature set can easily reach a size close to \textasciitilde1,000 features only when considering the original image intensities \cite{radiomics_1k_5}. If filtering is considered prior to feature extraction, the feature set size could rise to \textasciitilde10,000 and more \cite{radiomics_10k_6}. Overall, the variability in feature types requiring increasing computational steps gives rise to the concept of \textit{complexity of radiomic feature extraction}.

Furthermore, radiomics studies often integrate machine learning processes aiming at building predictive models from these extracted features with the goal, for example, to classify tumors and predict patient outcomes. The high dimensionality of radiomic feature sets introduces yet another level of complexity to the overall radiomic analysis pipeline, necessitating careful feature selection and model optimization to avoid overfitting and ensure generalizability \cite{how_to_7}. Consequently, more emphasis should be given to focusing on the most effective features for a given clinical question. For instance, in an overall survival prediction in lung cancer \cite{vulnerabilities_8}, it was found that the tumor volume alone was driving the prognostic accuracy and that intensity and texture values were not relevant for prognostication. On the other hand, more complex features such as gray–level histograms combined with texture features extracted from high-resolution CT images, following wavelet transforms, demonstrated a high accuracy in identifying lung tissue types \cite{depeursinge_lung_9}. This suggests that certain types of features are more suitable than others for predictive modeling in specific radiological clinical questions.

In this study, we propose a new methodology that identifies the most predictive features specific to a given clinical outcome and modality, and explains the model's choices, further enhancing understanding and potentially improving generalizability in the future. By analyzing five datasets, each with a distinct medical indication across two imaging modalities, we demonstrate how our methodology simplifies radiomics analysis by focusing on the most relevant features, and we also demonstrate the capabilities of MEDimage, an openly accessible tool designed to to implement our methodology and to potentially promote synergy between clinical radiologists and computer scientists through both a code-based solution and a user-friendly interface. Finally, we highlight how focusing on a single complexity level can improve performance.

\section{Materials and Methods}
\label{sec:mats_meths}

\subsection{Cohorts}
We collected five distinct imaging cohorts of cancer patients, including non-small cell lung cancer (NSCLC), low-grade glioma (LGG), meningioma and two imaging cohorts of renal cell carcinoma (RCC). The NSCLC dataset was collected from Primakov \textit{et al.} study \cite{nsclc_study}. It includes data from three institutions: MAASTRO \cite{nsclc_dataset_10}; Stanford \cite{nsclc_dataset_11}, available on The Cancer Imaging Archive (TCIA) \cite{TCIA_12}; And the University of California San Francisco (UCSF), which is not public. For LGG, part of the data was collected from The Cancer Genome Atlas (TCGA) \cite{lgg_tcga_13}, while the rest was provided by Yu Je \textit{et al.} and Li Z \textit{et al.} studies \cite{lgg_dataset_14, lgg_dataset_15}, respectively. Meningioma cohort was provided by Wu \textit{et al.} \cite{wu_presenting_2018} , Vasudevan \textit{et al.} \cite{vasudevan_comprehensive_2018}, Gennatas \textit{et al.} \cite{gennatas_preoperative_2018} and Morin \textit{et al.} \cite{meningioma_dataset_16} studies. MRI-based RCC cohort was provided by Lin Xi \textit{et al.} study \cite{rcc_dataset_17} , and TCGA data \cite{dataset_rcc_18} was included. Finally, the contrast-enhanced computed tomography (CECT)-based RCC dataset was provided by CIUSSSE-Centre hospitalier universitaire de Sherbrooke (CHUS). Institutional review approval was given for the use of this dataset in this retrospective study. The cohorts are summarized in table \ref{tab:dataset} (patient characteristics are provided in supplementary note 3).

\begin{table}[ht]
    \scriptsize
    \begin{tabular}{|llllll|}
    \hline
    \multicolumn{1}{|l|}{Cohorts}                                   & \multicolumn{1}{c|}{\begin{tabular}[c]{@{}c@{}}NSCLC\\ (n=506)\end{tabular}}                                                                            & \multicolumn{1}{c|}{\begin{tabular}[c]{@{}c@{}}LGG\\ (n=329)\end{tabular}}                                                        & \multicolumn{1}{c|}{\begin{tabular}[c]{@{}c@{}}Meningioma\\ (n=344)\end{tabular}}                                                                            & \multicolumn{1}{c|}{\begin{tabular}[c]{@{}c@{}}RCC\\ (n=599)\end{tabular}}                                                                                                       & \multicolumn{1}{c|}{\begin{tabular}[c]{@{}c@{}}RCC\\ (n=326)\end{tabular}}                                                             \\ \hline
    \multicolumn{1}{|l|}{Institutions}                              & \multicolumn{1}{l|}{\begin{tabular}[c]{@{}l@{}}MAASTRO (n=207)\cite{nsclc_study, nsclc_dataset_10}\\ UCSF (n=163) \cite{nsclc_study}\\ Stanford (n=136)\cite{nsclc_study, nsclc_dataset_11}\end{tabular}}                                & \multicolumn{1}{l|}{\begin{tabular}[c]{@{}l@{}}TCGA (n=103)\cite{lgg_tcga_13}\\ Huashan (n=226)\cite{lgg_dataset_14}\\ \cite{lgg_dataset_15}\end{tabular}}                          & \multicolumn{1}{l|}{\begin{tabular}[c]{@{}l@{}}UCSF (n=257)\cite{wu_presenting_2018}\\\cite{vasudevan_comprehensive_2018, gennatas_preoperative_2018, meningioma_dataset_16}\\ PM (n=87)\cite{meningioma_dataset_16}\end{tabular}}                                                              & \multicolumn{1}{l|}{\begin{tabular}[c]{@{}l@{}}Penn (n=439)\cite{rcc_dataset_17}\\ Mayo (n=53) \cite{rcc_dataset_17}\\ TCGA (n=54)\cite{dataset_rcc_18}\\ HPH (n=30)\cite{rcc_dataset_17}\\ XYSH (n=23)\cite{rcc_dataset_17}\end{tabular}}                        & \begin{tabular}[c]{@{}l@{}}CIUSSSE-CHUS\\ (n=326)\end{tabular}                                                                         \\ \hline
    \multicolumn{1}{|l|}{Imaging modality}                          & \multicolumn{1}{c|}{CT}                                                                                                                                 & \multicolumn{1}{c|}{MRI-T2F}                                                                                                      & \multicolumn{1}{c|}{MRI-T1CE}                                                                                                                                & \multicolumn{1}{c|}{MRI-T2WI}                                                                                                                                                    & \multicolumn{1}{c|}{CECT}                                                                                                              \\ \hline
    \multicolumn{1}{|l|}{Clinical endpoint}                         & \multicolumn{1}{l|}{\begin{tabular}[c]{@{}l@{}}Histological subtype:\\ - Adenocarcinoma (n=240)\\ - Other (n=266)\end{tabular}}                         & \multicolumn{1}{l|}{\begin{tabular}[c]{@{}l@{}}IDH1 mutation:\\ - Yes (n=239)\\ - No (n=90)\end{tabular}}                         & \multicolumn{1}{l|}{\begin{tabular}[c]{@{}l@{}}Pathological tumor \\grade*:\\ - Grade 1 (n=197)\\ - Grade 2 \& 3 (n=147)\end{tabular}}                         & \multicolumn{1}{l|}{\begin{tabular}[c]{@{}l@{}}Subtype discrimination:\\ - Papillary (n=158)\\ - Clear Cell (n=441)\end{tabular}}                                                & \begin{tabular}[c]{@{}l@{}}Subtype discrimination\\ - Non-Clear Cell (n=79)\\ - Clear Cell (n=247)\end{tabular}                        \\ \hline
    \multicolumn{6}{|l|}{\begin{tabular}[c]{@{}l@{}}CT: Computed tomography. MRI: Magnetic resonance imaging. T2F: T2-weighted/FLAIR (Fluid attenuated inversion recovery).\\T1CE: T1-weighted contrast-enhanced. T2WI: T2-weighted image. \\ CECT: Contrast-enhanced computed tomography. UCSF: University California San Francisco. PM: Princess Margaret Cancer Centre. \\ CIUSSSE-CHUS: Centre intégré universitaire de santé et de services sociaux de l’Estrie-Centre hospitalier universitaire de Sherbrooke. \\ Penn: Hospital of the University of Pennsylvania. Mayo: Mayo Clinic. HPH: Hunan Provincial People's Hospital. \\ TCGA: The Cancer Genome Atlas. XYSH: Xiangya Second Hospital of Central South University. \\ *For binary prediction of pathological grade, Grade 1 is considered “Low” (0), and Grade 2 and 3 is considered “High” (1).\end{tabular}} \\ \hline
    \end{tabular}
    \caption{Study cohorts summary}
    \label{tab:dataset}
    \end{table}

\subsection{Radiomic features and levels}
For each dataset, a total of 170 features were extracted from the original images, in accordance with IBSI definitions \cite{ibsi_1}: 27 shape-based; 49 intensity-based (1 local intensity, 18 statistical, 23 intensity histogram and 7 intensity-volume histogram); and 94 texture-based, each extracted using six different combinations of parameters, for images with physical units (CT and CECT) and three different combinations of parameters for the rest (MRI and filtered images). For filtered images, shape and intensity-based statistical features were excluded as they are not meaningful for arbitrary intensity scales, resulting in a total of 124 features. Six LF were used, including mean, Laplacian-of-Gaussian, Laws, Gabor, wavelets (high-pass and low-pass), all in accordance with IBSI definitions \cite{ibsi_2}. We also used nonlinear filtering by moving a predefined 3D cubic window over the voxels of the image and calculating texture feature values from the Gray Level Co-occurrence Matrix (GLCM) family at each position (hereby referred to as “textural filtering”). This resulted in 25 textural filters (TF) corresponding to the 25 features defined in the GLCM family. To our knowledge, no reference values currently exist for such filters, but similar definitions are found in the works of Mayerhoefer \textit{et al.} \cite{intro_radiomics_19}, and of Deasy \textit{et al.} \cite{cerr_20}. Consequently, a total of 18,112 features were extracted for cohorts with physical units, and 17,830 features for cohorts with arbitrary ones. Image processing and extraction settings are provided in supplementary note 1.

The aforementioned categories of radiomic features are henceforth designated as complexity levels. Therefore, our investigation focuses on five levels of complexity. The first three levels include features extracted from the original image intensities: Morphological (“M”) features, Intensity-based (“I”) features and Texture-based (“T”) features. The fourth (“LF”) and fifth (“TF”) levels are features extracted after linear and textural filtering, respectively. To increase complexity, features were combined prior to predictive modeling, giving rise to the final sequence of complexity levels: “M”, “I”, “M+I” (MI), “T”, “M+I+T” (MIT), “LF”, “M+I+T+LF” (MITLF), “TF”, and “M+I+T+LF+TF” (MITLFTF).

\subsection{MEDimage}
To facilitate synergy between clinical radiologists and computer scientists, our approach was built on two integral components. Firstly, we have developed a Python-based package with a modular architecture, ensuring flexibility of the code. Each module is dedicated to specific tasks in feature extraction and model training. Secondly, we have implemented a node-based user interface (UI) based on Electron and ReactFlow, offering clinical radiologists easy access to various modules for model training, testing, and results analysis, without requiring programming skills. Importantly, users can transition between the two components by automatically generating Python code for selected experiments on the interface. A comparison of existing free-to-access radiomics tools is available in the supplementary table 2. All information about the software is available here: \url{https://medimage.app}.

\subsection{Experiment workflow}
The experiment design was separated into three phases. The initial phase involved processing raw data and extracting radiomic features. Features were then organized by complexity levels, starting from “M” to “MITLFTF”. Following this, 10-fold cross-validation was used to partition the data. At each complexity level and for every fold, features were reduced using an adaptation of the false-discovery-avoidance method \cite{fda_21}, retaining only a small subset (n\textasciitilde5-20) with the least intra-correlation and highest correlation to the outcome. Subsequently, models were then trained using the XGBoost algorithm, a gradient-boosted decision tree, across four feature counts (5, 10, 15, and 20 most relevant) to assess whether increasing the number of features enhances performance. Testing folds were utilized to evaluate model performance. Finally, the analysis of results was based on feature importance, which quantifies the improvement in performance brought by a feature during the model's construction. The analysis involved two key steps: (i) Identification: A heatmap of metrics was used to pinpoint the optimal complexity level, characterized by the minimum number of features, minimum complexity, and the most statistically significant performance; and (ii) Explanation: For the selected level, feature importance tree, a novel method, was utilized. It consists of a tree plot that breaks down the selected complexity level in a cascade architecture, where each branch is connected to the filter, feature family and individual features that contributed to the decision-making process. Branch thickness reflects the feature importance, and the path that leads to the most predictive individual feature is highlighted. Additionally, a feature importance histogram was employed to display the importance scores of features, to assess contribution to the model's predictive performance. The workflow is illustrated in Figure \ref{fig:exp_workflow} with a detailed version in supplementary note 2.

\begin{figure}
  \centering
  \includegraphics[scale=0.33]{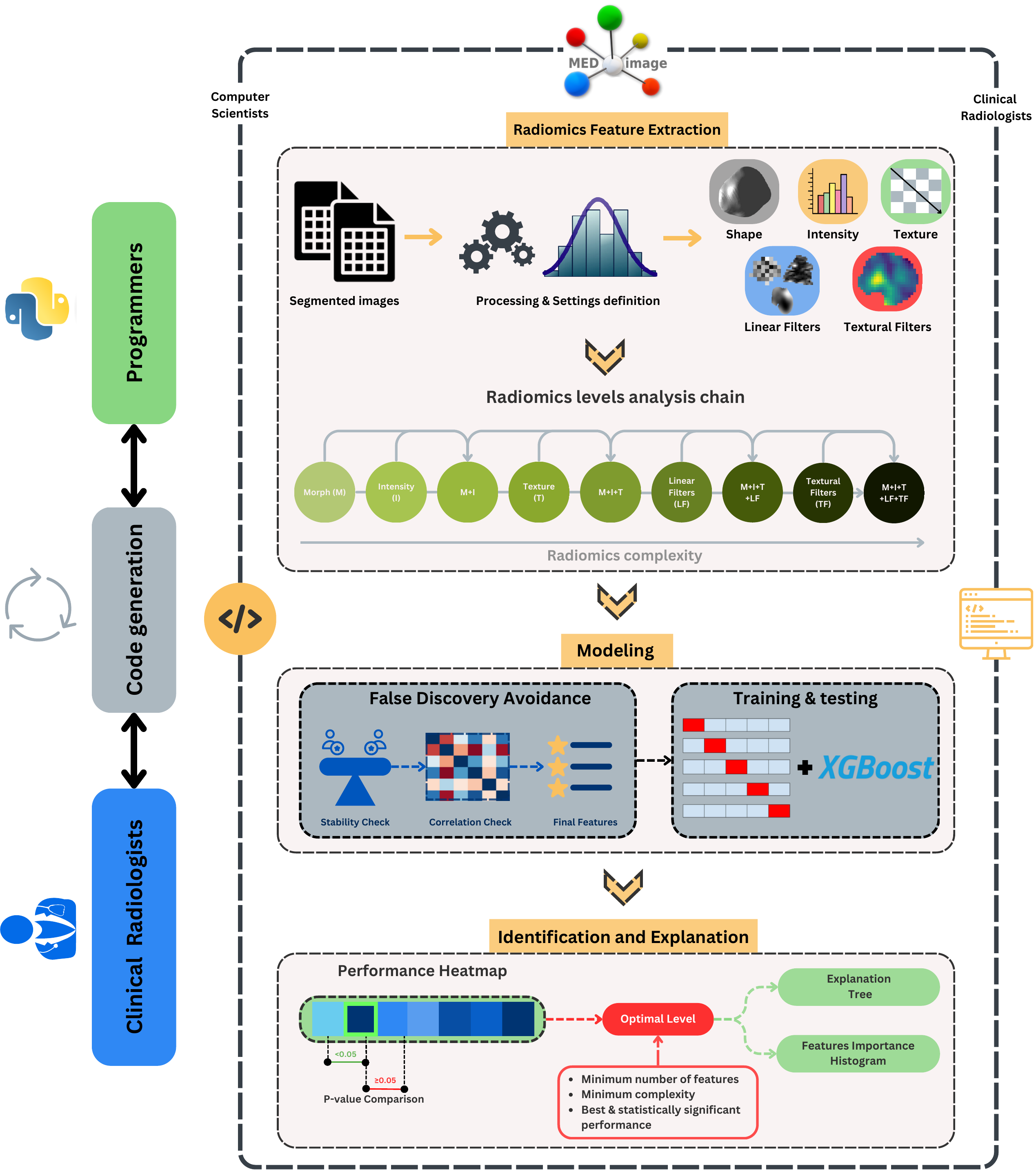}
  \caption{Overview of the study workflow. The workflow in a typical radiomics analysis starts with acquisition and reconstruction of medical images. Subsequently, images are segmented to define regions of interest (ROIs). Following this step, the proposed radiomics software processes the images and computes features characterizing the ROIs, which are then organized by complexity levels for model training. Machine learning begins with feature cleaning to remove or replace invariant features, followed by feature set reduction to retain features exhibiting a high and stable correlation with the clinical endpoint, while removing inter-correlated features. All models are constructed using XGBoost. The final step involves results analysis through two stages: identification of the optimal complexity level, characterized by: the minimum number of features; minimum complexity; the highest and statistically significant performance, and explanation based on feature importance. Experiments can be conducted via programming or through the interface, with the code generation option facilitating the shift between the two approaches.}
  \label{fig:exp_workflow}
\end{figure}

\subsection{Statistical analysis}
The area under the curve (AUC) metric was employed for evaluation. To address the issues encountered by small datasets, including overfitting, noise, outliers and sampling bias, which can render the learned model ineffective \cite{issue_small_data_22}, model predictions over all 10 test folds of the cross-validation were aggregated into a single receiver operating characteristic (ROC) curve, mimicking the behavior of leave-one-out cross-validation. While this approach sacrifices the model’s variance, it is known for minimizing bias and offering reliable estimates \cite{cv_review_23}. A DeLong test \cite{delong_24} was used to determine whether models were statistically different, and the difference was considered significant with $\text{p-value}<.05$.

\section{Results}
\subsection{MEDimage}
We have developed MEDimage, an open-source tool designed to streamline radiomics studies. Clinical radiologists can customize radiomics studies graphically by defining data processing and feature extraction sequences, through node moving and linking (showcased in supplementary media), facilitating multiple hypothesis testing and results comparison. For programmers, the modular implementation of the code ensures its easy manipulation. With the code generation option, users can transition from the graphical to the code-based approach.

\subsection{Optimal complexity level identification and explanation}
\begin{figure}
  \centering
  \includegraphics[scale=0.172]{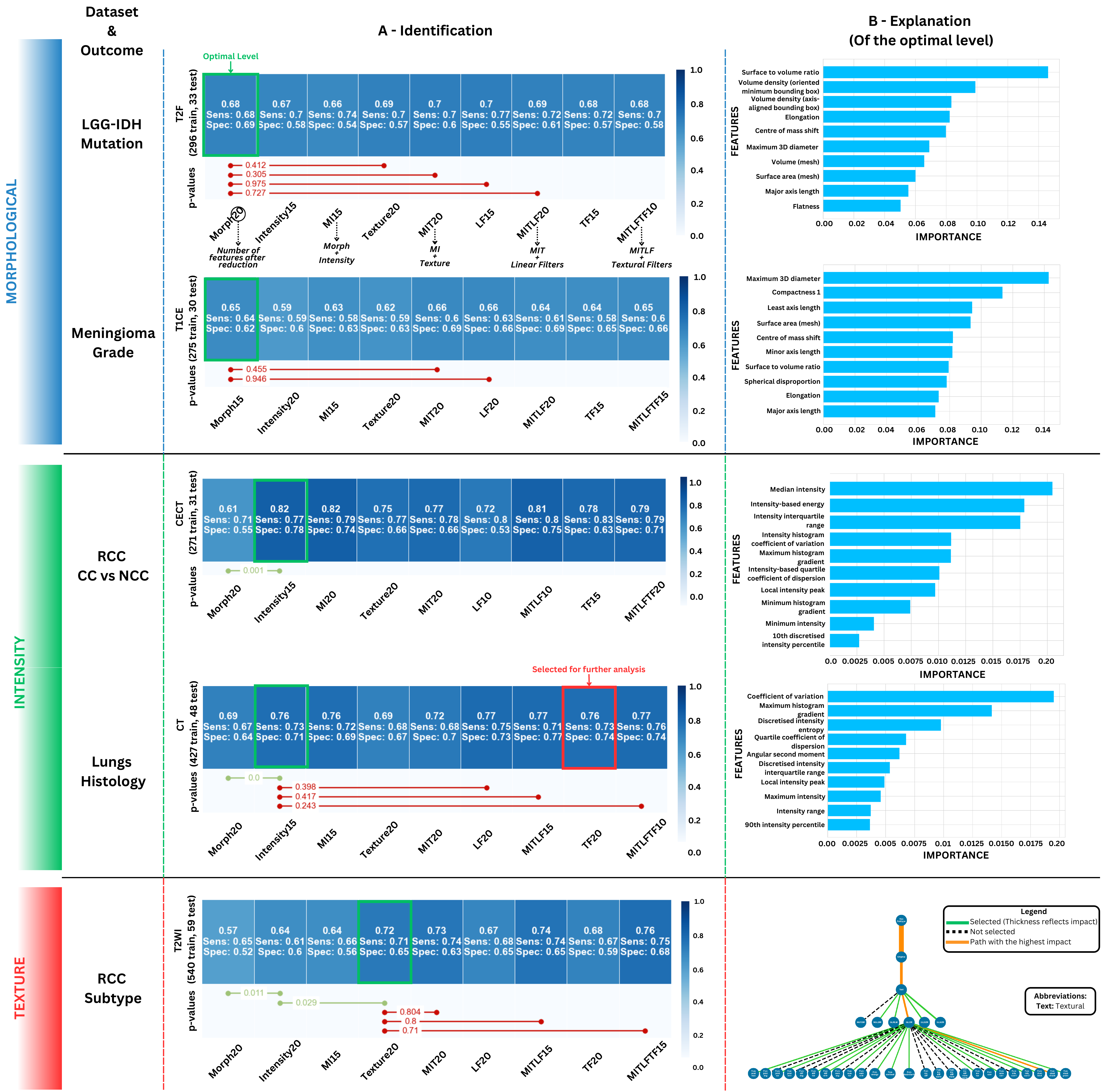}
  \caption{Results overview per optimal complexity level and per cohort. Levels are identified as blue (morphological), green (intensity), and red (texture). A. Comparison of heatmaps showing aggregated AUC across splits for various radiomics signatures. Each column indicates a complexity level. The numerical suffixes in column names indicate the number of features retained after feature set reduction. In the bottom rows, green lines indicate significant improvements ($\text{p-value}<.05$) and red lines indicate non-significant improvements. Green boxes point to the identified optimal levels and the red one is selected for a further discussion in supplementary material. B: Explanation section displaying histogram of feature importance or importance tree, highlighting features with high importance at the identified optimal level.}
  \label{fig:results}
\end{figure}

Models performance was assessed using the aggregated AUC value across cross-validation splits as a base metric, and the mean feature importance across splits was used to identify highly predictive features. All results are depicted in Figure \ref{fig:results}.

\subsubsection{Morphological features as the optimal complexity level:}
In the Meningioma cohort, the model solely utilizing morphological features displayed an AUC of 0.65 (95\% CI: 0.59, 0.72), specificity of 0.62, and sensitivity of 0.64. Models based on other complexity levels displayed varying degrees of performance, such as intensity-based (AUC 0.59), MI-based (AUC 0.63), Texture-based (AUC 0.62), and TF-based (AUC 0.64). MIT-based and MITLF-based models had the highest specificity (0.69) but a lower sensitivity (0.6 and 0.61, respectively). MIT-based and LF-based models exhibited the highest AUC (0.66), yet were not different from the morphological features-based model (.46 and .95 respectively). Therefore, the morphological features based model was selected as optimal. The maximum 3D diameter feature had the highest mean importance in the model. Similarly, for the prediction of IDH1 mutation in LGG, morphological features were sufficient to obtain the best performance, with an AUC of 0.68 (95\% CI: 0.60, 0.75), specificity of 0.69, and sensitivity of 0.68. The surface to volume ratio feature had the highest mean importance.

\subsubsection{Intensity features as the optimal complexity level:}
For histological subtype classification of NSCLC, the intensity features model achieved an AUC of 0.76 (95\% CI: 0.71, 0.80), specificity of 0.71, and sensitivity of 0.73. Models based on LF, MITLF and MITLFTF demonstrated higher AUC values (0.77), yet their p-values (.40, .42 and .24, respectively) did not indicate statistical significance with respect to the intensity features model. The coefficient of variation had the highest mean feature importance. Similarly, for the clear and non-clear RCC classification based on CECT, the model based on intensity features recorded the highest AUC of 0.82 (95\% CI: 0.76, 0.88), with a specificity of 0.78 and a sensitivity of 0.77. Median intensity had the highest mean feature importance.

\subsubsection{Texture features as the optimal complexity level:}
Texture-based model proved improvement in performance from morphological-based (p=.01) and intensity-based (p=.03) models, for the subtype classification of RCC based on MRI-T2WI, achieving an AUC of 0.72 (95\% CI: 0.68, 0.77), specificity of 0.65, and sensitivity of 0.71. Within the texture feature families, the GLCM feature family demonstrated the highest mean feature importance, with cluster shade having the highest importance.

\subsubsection{Linear filters and textural filters:}
Although the LF-based models were never selected as optimal in any case, they consistently demonstrated high AUC values across cohorts. For example, in the classification of NSCLC subtypes, the LF-based model achieved an AUC of 0.77 (95\% CI: 0.72, 0.80), indicating their potential utility in radiomics analyses.

Similarly, TF-based models, though not selected as optimal in any cohort, demonstrated high AUC values across various cohorts. For instance, in NSCLC subtype classification, the TF-based model matched the AUC of the selected optimal level (0.76; 95\% CI: 0.72, 0.80), but was not considered optimal due to its higher complexity (detailed results provided in supplementary note 5). Figure \ref{fig:tf} illustrates the application of a textural filter on NSCLC images, selected from patients with the highest difference in the feature with the highest importance.

\begin{figure}
  \centering
  \includegraphics[scale=0.4]{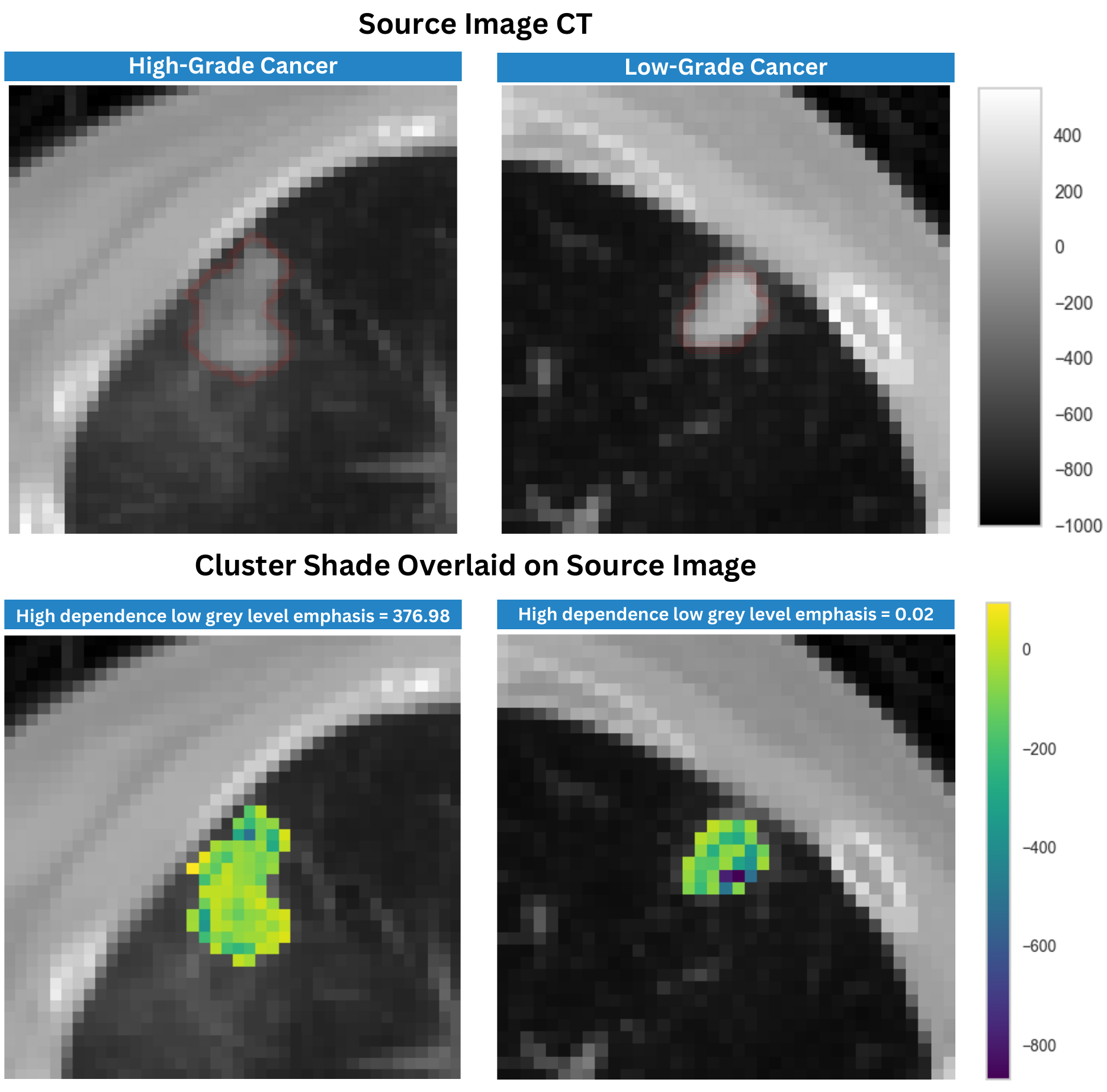}
  \caption{CT images of low-grade (right) and high-grade (left) NSCLC showing the application of cluster shade texture analysis. Images were selected based on the highest difference in high dependence low grey level emphasis (hdlge) feature (376.98 for high-grade and 0.02 for low-grade cancer), which had the highest importance after training. A contrast difference is visible between low and high-grade cancers (top). Overlay of the cluster shade revealed intratumoral heterogeneity (bottom).}
  \label{fig:tf}
\end{figure}

\subsection{In-depth analysis of an optimal complexity level}
For clear and non-clear RCC classification, the intensity-based level was selected as optimal, and the median intensity had the highest feature importance. To emphasize its impact, images were automatically selected from patients with the highest difference in the median intensity measure and were displayed to visually assess the distinctions between clear and non-clear cell carcinoma based on CECT (See Fig. \ref{fig:analyse_app}.A) (For other cohorts, comparisons are available in supplementary note 4). The clear-cell subtype typically exhibited hypervascularity and greater heterogeneity due to necrotic areas compared to the non-clear cell subtype. Moreover, Identifying the optimal level allowed us to refine feature extraction settings, particularly the re-segmentation range \cite{ibsi_manu_1_25}, which directly affects the intensities inside the ROI \cite{ct_range_impact}. This refinement led to a 4\% improvement in AUC from 0.82 to 0.86 (See Fig. \ref{fig:analyse_app}.B).

\begin{figure}
  \centering
  \includegraphics[scale=0.5]{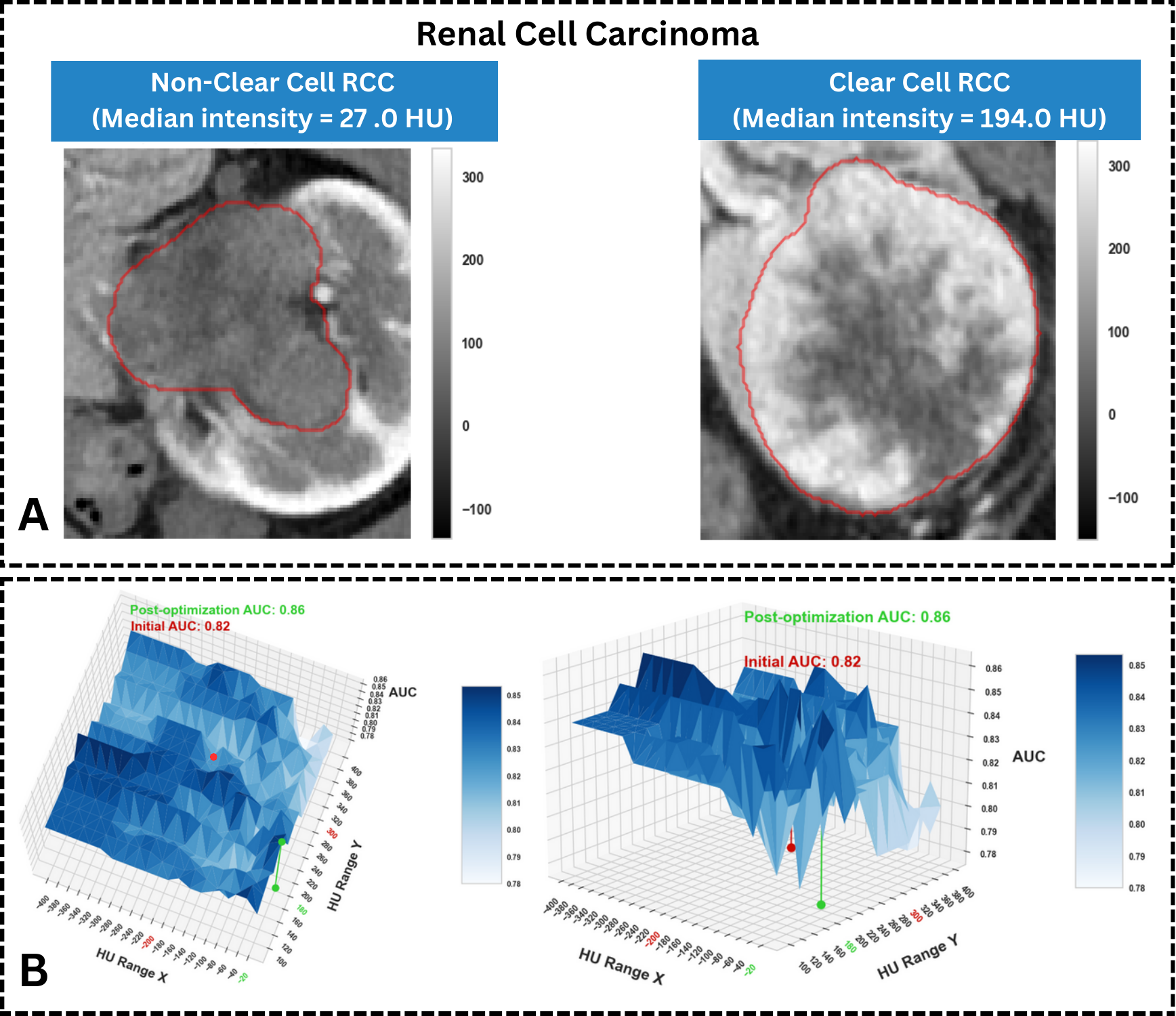}
  \caption{(A) Axial contrast enhanced computed tomography (CECT) comparison of a clear cell versus non-clear cell renal carcinoma with the highest disparity in median intensity. (B) Different views of 3D heatmap illustrating the influence of Hounsfield Unit (HU) range $[X,Y]$ on machine learning results in clear cell and non-clear cell renal carcinoma classification. Range limits highlighted in red denote the pre-optimization limits, while those in green signify the updated limits post-optimization.}
  \label{fig:analyse_app}
\end{figure}

\section{Discussion}
Many studies have highlighted the potential of radiomics to enhance clinical decision making, but application requires further optimization and standardization \cite{how_to_7}, In this work, the methodology we developed aims to simplify radiomics predictive modeling, paving the way for future clinical applications. We introduced the concept of radiomics complexity levels defined by the number of computational steps needed to extract features, and proposed a methodology for estimating an optimal radiomics complexity level for a given clinical problem, that takes into account computational steps, predictive performance, and statistical significance, in order to focus on predictive features and potentially pave the way for more generalizability. Additionally, we proposed MEDimage, an innovative software tool designed to streamline radiomics studies, and facilitate synergy between computer scientists and clinical radiologists.

Selections of optimal levels aligned with findings from other available studies, indicating the robustness of our methodology. For example, shape features had the highest impact in meningioma cancer grade classification which corroborates the findings of Zhang \textit{et al.} \cite{zhang_26} who reported a correlation between shape features and brain invasion in meningioma cancer grade prediction. Similarly, texture features were found predictive and sufficient in MRI-based non-clear cell and clear cell RCC classification, corroborating the findings of Wang \textit{et al.} \cite{wang_rcc_27}, who found texture features effective in differentiating three RCC subtypes (clear cell, papillary, and chromophobe) from MRI images. Finally, Linning \textit{et al.} \cite{lungs_linning_28} utilized radiomics for classifying histological subtypes of lung cancer and found intensity features to be the most predictive, indicating tumor heterogeneity. These findings are concordant with our results and suggest our methodology could pave the way for more generalizability across diverse clinical scenarios.

All experiments conducted as part of our study utilized MEDimage, offering enhanced flexibility in study design and analysis. It facilitates feature extraction, cleaning, selection, model training, and results analysis. Feature selection is particularly relevant for radiomics approaches to choose predictive biomarkers. Thus, our selection involved applying standard false-discovery-avoidance method \cite{fda_21} independently to each feature type, followed by a final iteration on the combination of all types. We also accounted for variants of texture features extracted under different parameter sets, enhancing robustness of our estimations against changes in extraction settings related to texture features selection. In results analysis, the feature importance tree drew the path that highlighted filters, feature families and individual features that significantly contributed to model performance, offering insights into the model's decision-making process and enabling further optimization.

Our aim in developing the workflow and software outlined here was to streamline the process by avoiding unnecessary complexity and emphasizing the efficacy of less complex features for optimal performance, potentially increasing generalizability. Through this approach, we believe that the focus on a singular optimal level can potentially save time by preventing the investigation of irrelevant features, while also laying the groundwork for a more in-depth analysis. For example, in the case of clear cell versus non-clear cell RCC classification, 18,112 features were extracted and tested, however, we later identified that a set of 15 intensity-based features was sufficient to obtain the best performance. We then optimized the Hounsfield unit range used in the ROI re-segmentation step \cite{ibsi_manu_1_25}, which directly affected the intensities inside the ROI, and improved the AUC from 0.82 to 0.86.

Our study has some limitations. First, access to large datasets restricted our ability to assess generalizability and study cases that could benefit from different optimal levels such as nonlinear filters. Additionally, other combinations of complexity levels, such as combining morphological and texture features, were not explored, potentially missing out on improvements in predictive performance. Our estimation of optimal complexity levels is susceptible to variations in image processing and feature extraction parameters. We exclusively used the GLCM texture family for nonlinear filtering, suggesting future exploration of other textural feature families. Also, we did not assess the robustness of features against differences in positioning, acquisition and segmentation \cite{zwanenburg_robustness_29}, potentially leaving the identified optimal levels susceptible to differences in these factors. Additionally, all classifications were limited to binary problems and exclusively analyzed using the XGBoost algorithm. Inclusion of features derived from deep learning represents an area for future investigation, adding an additional layer of complexity to radiomic analyses. These limitations underscore the need for future exploration into more automated and robust techniques for selecting optimal levels to enhance the efficacy and reproducibility of radiomics approaches.

To conclude, our study unveiled context-specific optimal radiomics complexity levels, as demonstrated across five distinct datasets. Leveraging our proposed methodology and software, we successfully identified and explained the optimal level for each dataset, providing an optimal simplification of radiomics use in predictive modeling.

\section*{Acknowledgments}
We thank Nicolas Longchamps, BA, Guillaume Blain, BA, Sarah Denis, MS and Andréanne Allaire, BA, for their valuable contribution to the development of the software’s user interface.

\section*{Author contributions}
Guarantors of integrity of entire study, M.A.L.L., J.S., S.L., O.M., C.R., M.L., M.V.; study concepts/study design or data acquisition or data analysis/interpretation, all authors; manuscript drafting or manuscript revision for important intellectual content, all authors; approval of final version of submitted manuscript, all authors; agrees to ensure any questions related to the work are appropriately resolved, all authors; literature research, M.A.L.L., C.R., M.L., M.V.; clinical studies, T.B.P., D.R.R., W.C.C., D.P.I.C., H.Z., O.G., J.W., A.C.S., P.J.Z., H.X.B., P.O.R., O.M.; experimental studies, M.A.L.L.; statistical analysis, M.A.L.L.; code contribution, A.Z., T.U., J.B.; and manuscript editing, M.A.L.L., O.M., C.R., M.L., M.V.

\section*{Funding information}
Martin Vallières acknowledges the support of the Canada CIFAR AI Chairs Program and Unité de Soutien SSA QC. Harrison X. Bai acknowledges funding from the National Institute of Health (NIH) under Project \#1R03CA249554-01. David R Raleigh acknowledges funding from the NIH under Project R01 CA262311. Jan Seuntjens acknowledges funding from the Canadian Institutes of Health Research (FDN-143257). 

\section*{Data sharing statement}

At this time, the following datasets are not publicly shared by the hosting institutions: Non-Small Cell Lung Cancer (UCSF), Meningioma (UCSF, PMH), Renal cell carcinoma 1 (Penn, Mayo, HPH, XYSH) and Renal cell carcinoma 2 (CHUS). The Non-Small Cell Lung Cancer data from MAASTRO is available here: \url{https://doi.org/10.7937/K9/TCIA.2015.PF0M9REI} and data from Stanford is available here: \url{https://doi.org/10.7937/K9/TCIA.2017.7hs46erv}. The Low-grade glioma dataset from Huashan is available upon reasonable request here: \url{https://doi.org/10.1038/s41598-017-05848-2}; \url{https://doi.org/10.1007/s00330-016-4653-3}. The Low-grade glioma dataset from TCGA is available here: \url{http://doi.org/10.7937/K9/TCIA.2016.L4LTD3TK}. The Renal cell carcinoma dataset from TCGA is available here: \url{https://doi.org/10.7937/K9/TCIA.2016.V6PBVTDR}.

\begin{center}
    \Huge\textbf{Supplementary Notes}
\end{center}

\section*{Supplementary Note 1: Image processing configurations}
\label{sec:supp_image_processing}

    The image processing configurations used prior to feature extraction for each cohort are listed in table \ref{tbl:configurations}.
    
    \begin{table}[htbp]
        \centering
        \footnotesize
        \begin{tabular}{|c|m{1.75cm}|m{1.75cm}|m{1.75cm}|m{1.75cm}|m{1.75cm}|m{1.75cm}|}
        \hline
        \multicolumn{2}{|c|}{} &\multicolumn{5}{c|}{Image processing configurations} \\
         \cline{3-7}
         
         \multicolumn{2}{|c|}{} & Lungs Cancer (CT) & Low-grade glioma (MRI) & Meningioma (MRI) & Renal-cell \newline carcinoma (MRI) & Renal-cell \newline carcinoma (CECT)\\ \hline
        
        \multirow{5}{*}{\rotatebox{90}{Interpolation \hspace{0.5cm} }} & Voxel dimension (mm) & 2×2×2 & 1×1×1 &  1×1×1 & 4x4x4 & 2×2×2 \\\cline{2-7}    
        & Interpolation method & Tricubic spline &  Tricubic spline &  Tricubic spline &  Tricubic spline &  Tricubic spline \\\cline{2-7}
       & Intensity rounding & Nearest integer & - &  - &  - &  Nearest integer \\\cline{2-7}
       & ROI interpolation method & Trilinear &   Trilinear & Trilinear & Trilinear & Trilinear \\ \cline{2-7}
       & ROI partial mask volume threshold & 0.5 & 0.5 & 0.5 & 0.5 &   0.5\\\hline

       \multirow{3}{*}{\rotatebox{90}{Discretization \hspace{0.2cm}}} & Intensity histogram & FBN: 64 bins & FBN: 64 bins &  FBN: 64 bins & FBN: 64 bins & FBN: 64 bins \\\cline{2-7}    
        & Intensity volume histogram & - & FBN: 1000 bins & FBN: 1000 bins &  FBN: 1000 bins &  - \\\cline{2-7}
       & Texture & FBN: [8, 16, 32] bins; FBS: [31, 63, 125] HU & FBN: [8, 16, 32] bins &  FBN: [8, 16, 32] bins &  FBN: [8, 16, 32] bins &  FBN: [8, 16, 32] bins; FBS: [16, 31, 63] HU \\\hline

       \multirow{2}{*}{Re-seg} & Intensity range & \(\left[-700, 300\right]\) HU &  \(\left[0, +\infty\right[\) &   \(\left[0, +\infty\right[\) &  \(\left[0, +\infty\right[\) &  \(\left[-200, 300\right]\) HU \\\cline{2-7}
        & Outlier filtering & - & Collewet* &  Collewet &  Collewet &  - \\\hline

       \multirow{1}{*}{Mean filter} & Size & 3 &  7 &  3 & 3 & 3 \\\hline

       \multirow{1}{*}{LoG} & Sigma (mm) & 1.75 &  4.2 &  2.16 & 1.5 & 2.23 \\\hline

       \multirow{5}{*}{Gabor} & Sigma (mm) & 1.75 &  4.2 &  2.16 & 1.5 & 2.23 \\\cline{2-7} 
        & Lambda (mm) & 3.5 &  8.4 & 8.4 & 3 & 4.45 \\\cline{2-7}
        & Theta & $\frac{\pi}{8}$ &  $\frac{\pi}{8}$ &  $\frac{\pi}{8}$ & $\frac{\pi}{8}$ & $\frac{\pi}{8}$ \\\cline{2-7}
        & Gamma & 2 &  2 & 2 & 2 & 2 \\\cline{2-7}
        & Rotation \newline invariance & yes &  yes &  yes & yes & yes \\\hline

        \multirow{1}{*}{Laws} & Kernel & [L3, L3, L3] &  [L5, L5, L5] & [L3, L3, L3] & [L3, L3, L3] & [L3, L3, L3] \\\hline

        \multirow{3}{*}{Wavelet} & Basis function & Coiflet 1 &  Coiflet 3 &  Coiflet 3 & Coiflet 1 & Coiflet 2 \\\cline{2-7} 
        & Subband & HHH, LLL & HHH, LLL & HHH, LLL & HHH, LLL & HHH, LLL \\\cline{2-7}
        & Level & 1 &  1 &  1 & 1 & 1 \\\hline

        \multirow{3}{*}{Textural filtering} & Size & 7 &  7 &  3 &  3 & 3 \\\cline{2-7} 
        & Local discretization & FBN: 25 HU adapted* & FBN: 8 bins & FBN: 8 bins & FBN: 8 bins & FBN: 25 HU adapted* \\\cline{2-7}
        & Global discretization & FBN: 25 HU &  FBN: 8 bins &  FBN: 8 bins & FBN: 8 bins & FBN: 25 HU \\\hline

        \multicolumn{2}{|c|}{Boundary condition} & mirror padding &  mirror padding &  mirror padding & mirror padding & mirror padding \\\hline
       
        \end{tabular}
        \caption[Image processing parameters used for each cohort]{Image processing parameters used for each dataset. CECT: contrast-enhanced computed tomography; CT: computed tomography; ROI: region of interest; HU: Hounsfield Unit. FBN: Fixed bin number; FBS: Fixed bin size; H: High-pass; L: Low-pass; *ROI voxels outlier intensities were removed from the intensity mask using the method suggested by Collewet \textit{et al.} \cite{collewet}; **Adapted indicates that the bin number was computed using the specified bin width and the image intensity range.}
        \label{tbl:configurations}
    \end{table}

\section*{Supplementary Note 2: The MEDimage software}
\label{sec:supp_medimage}
    The MEDimage package (\url{https://github.com/MEDomics-UdeS/MEDimage}) can extract radiomic features from medical images using a modular implementation. In a typical workflow, the selected dataset undergoes automatic preprocessing before feature extraction. Subsequently, features are extracted according to user-defined parameters, with available tools aiding in selection of such parameters. Features are then extracted, predictive features are identified using an adaptation of the FDA method \cite{fda_21}, and models are trained and fine-tuned using the established machine learning library PyCaret (\url{https://pycaret.org/}). The platform facilitates the determination of optimal feature types for further analysis, and users can easily generate Python code for their experiments, promoting collaboration between clinical radiologists and computer scientists. MEDimage complies with international standards \cite{ibsi_1, ibsi_2} and provides comprehensive support through tutorials, videos, and detailed documentation. The following section highlights the major functionalities of the package and introduces the different modules used:
    
    \begin{itemize}
        \item Image pre-processing: Through the \textit{DataManager} class, MEDimage facilitates Digital Imaging and COmmunications in Medicine (DICOM) image management, including ROI management (reading, association with imaging volume, etc.). MEDimage reads and serializes the images into byte streams. This serialization process is referred to as “\textit{pickling}” (\url{https://docs.python.org/3/library/pickle.html}). The objects hold all necessary imaging data needed for extraction including for example  the tumor mask. The objects are also used to organize extraction results leading to the simplification and the minimisation of the code.
        
        \item Image Processing: Consists of interpolation, re-segmentation and other processing methods. All these methods are implemented in the \textit{processing} module. Image filtering is also implemented according to IBSI standards \cite{ibsi_2}, offering a choice of several built-in filters such as Laws, Gabor, etc. This includes non linear filters that are defined as radiomic feature maps generated by moving a defined cubic window over the voxels of the image and calculating feature value at each position, while each feature map depicts a single radiomic feature. Moreover, the textural features can be computed in two different ways, in the first one, we discretize the image intensities inside the ROI locally, meaning at each position of the cubic window, whereas in the second one, the discretization is done globally on the whole region.
        
        \item Feature extraction: The package module \textit{biomarkers} handles all feature extraction related processes. It allows feature extraction from single scans and batch data. For batch extraction, \textit{BatchExtractor} class is used where a parameter file is used to customize the extraction by setting the different parameters such as filter sizes. This class also generates and organizes results automatically.
        
        \item Model training: Model training within the MEDimage module encompasses various methods supporting both training and evaluation of machine learning models. Additionally, it offers users a range of useful techniques for preprocessing radiomics features, including cleaning, normalization, and feature selection.Currently, the package exclusively supports training using XGBoost and binary classification tasks. The sequential steps for model training are outlined as follows:

        \begin{itemize}
            \item Cleaning: Involves the removal of features considered irrelevant for the analysis eliminating those with low variance and a high number of missing patients. Additionally, patients with a high number of missing features are excluded from the dataset during this step.
            
            \item Normalization: Performed on features using the ComBat method \cite{combat} as a preprocessing step to mitigate batch effects.

            \item Feature set reduction (FSR): Implemented using the false discovery avoidance method (FDA) introduced by Chatterjee \textit{et al.} \cite{fda_21}. As illustrated in figure \ref{fig:supp_fda_breakdwon}, this method involves subdividing training sets into 100 internal training and validation splits using a 2:1 ratio, and stratified random subsampling. Variables with low stability, measured by Spearman’s rank correlation with the outcome of interest ($RS_{f/o}$), are discarded using a minimal $RS_{f/o}$ cut-off of 0.5. Additionally, inter-correlated variable pairs are removed using a maximal $RS_{f/f}$ cut-off of 0.7, resulting in the retention of N features with the highest $RS_{f/o}$. The number of features to retain is set by the user. In our case, we tested four different numbers of retained features and kept the best performing one. Our package incorporates a balanced version of this method, which consists of applying FDA separately to each feature set to ensure consistency in the number of features drawn from each set. Subsequently, these drawn features are combined to form a final feature set, upon which FDA is reapplied. This process ensures that each set participates equally in the selection process, preventing the number of features in each set from affecting the overall reduction process.

            \begin{figure}[ht!]
                \centering
                \includegraphics[scale=0.18]{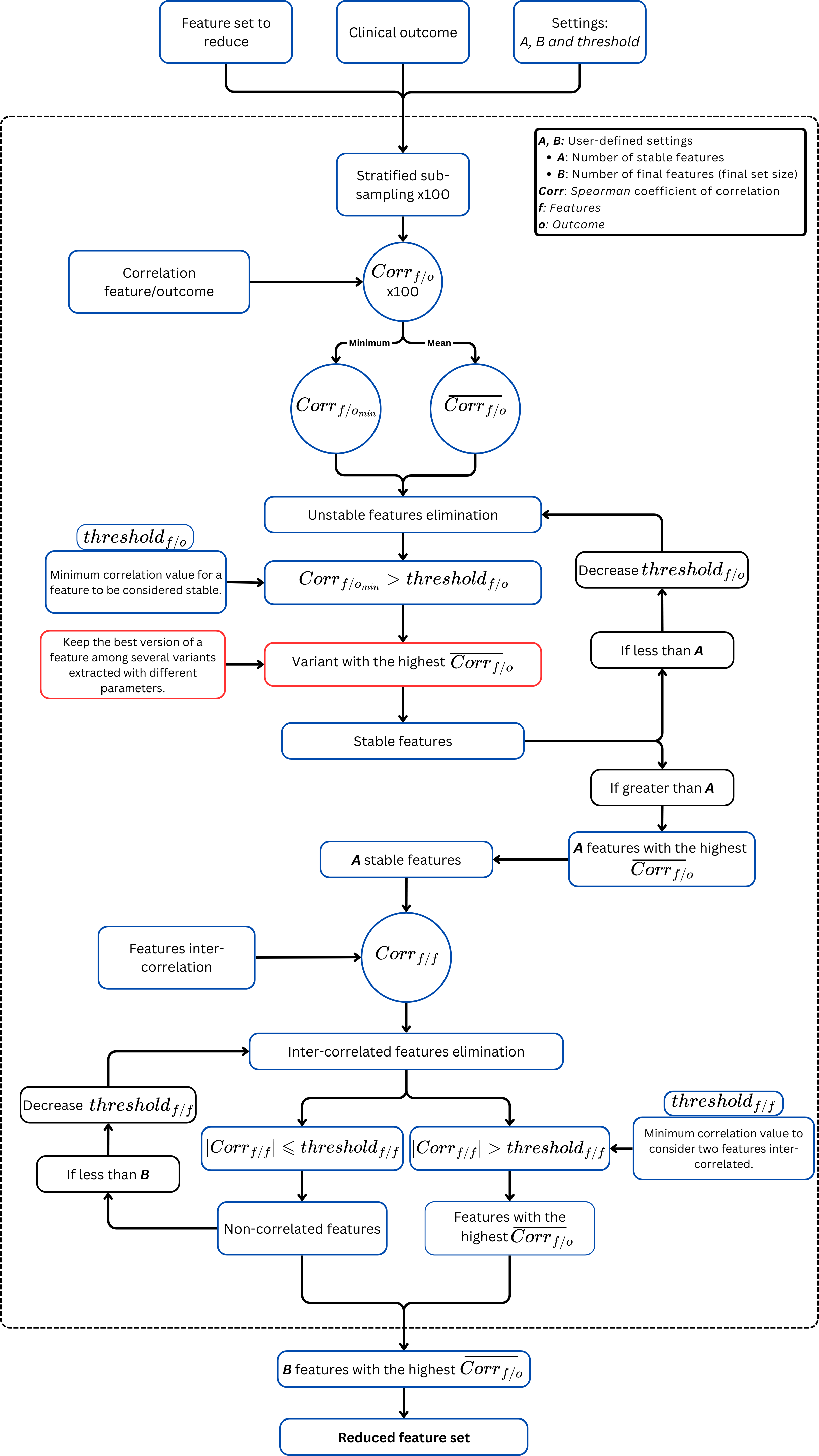}
                \caption[False discovery avoidance method breakdown]{False discovery avoidance method breakdown.}
                \label{fig:supp_fda_breakdwon}
            \end{figure}
        
            \item Model training: We relied exclusively on the XGBoost algorithm for binary prediction endpoints. The model was trained on the features retained after the feature set reduction step, with hyper-parameters automatically tuned using the pyCaret library. Subsequently, models were tested on the test set for all splits, with metrics computed for each split and aggregated or averaged to assess overall performance.
            
            \item Analysis method: Comprises two steps: identification and explanation of the optimal level. To identify the best-performing feature types, a heatmap is generated to compare the performance metric across complexity levels using p-values. The optimal level, characterized by the least number of features and the highest statistically significant performance metric, is then selected. In the explanation step, the histogram of feature importance is plotted, and for more complex cases (texture and filter-based features), the explanation tree highlights the feature family, filter, and features most impactful in the model's decision-making process.
        \end{itemize}
    \end{itemize}

    \subsection*{MEDimage app}
        To facilitate the utilization of the package by clinical radiologists, the MEDimage application (\url{https://github.com/MEDomics-UdeS/MEDimage-app}), also referred to as the MEDimage interface, provides access to package methods without requiring coding. It employs a drag-and-drop architecture where nodes are connected to form a pipeline representing the experiment, as depicted in supplementary figure \ref{supp_interface_pipelines}. Each node corresponds to a step in either feature extraction or model training, enabling customization of all processes and  facilitating multiple hypothesis testing and results comparison.
    
        \begin{figure}[h!]
            \centering
            \includegraphics[scale=0.30]{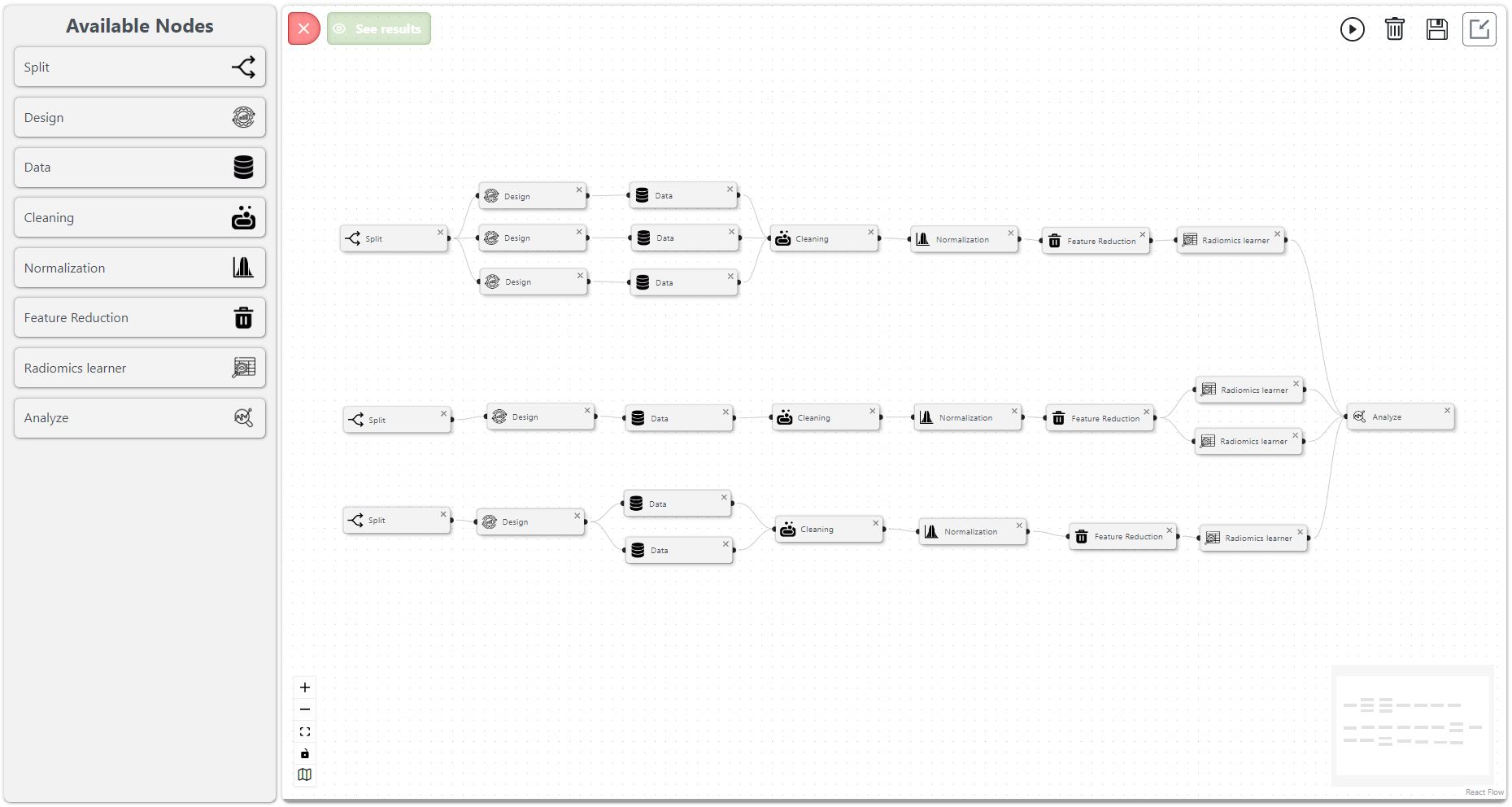}
            \caption[Illustration of multiple machine learning pipelines in the MEDimage interface.]{Illustration of multiple machine learning pipelines in the MEDimage interface.}
            \label{supp_interface_pipelines}
        \end{figure}

        The interface also supports results analysis for machine learning experiments, aligned with the methodology presented in this work. It facilitates easy comparison of metrics across training, testing, and holdout data, and provides convenient access to analysis plots, such as the feature importance histogram and the metrics heatmap, which includes statistical comparisons between models. This is illustrated in the supplementary figure \ref{supp_results_analysis}.

        \begin{figure}[h!]
            \centering
            \includegraphics[scale=0.70]{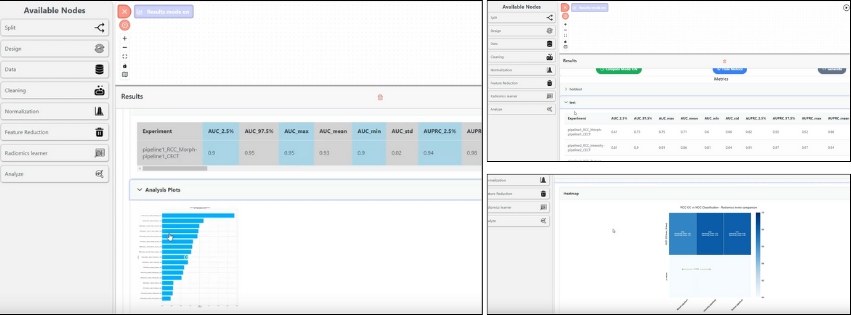}
            \caption[Analysis of Machine Learning Experiments Using the MEDimage App.]{Analysis of Machine Learning Experiments Using the MEDimage App. The left section displays results from a single experiment (pipeline) along with the corresponding analysis plot. The top right section compares metrics across experiments, and the bottom right section presents a metrics heatmap for the final models.}
            \label{supp_results_analysis}
        \end{figure}

        Finally, the MEDimage app supports code generation. Once the machine learning experiments are executed, users can select and generate code for multiple pipelines (experiments) through the interface, making the transition from a graphical interface to a code-based approach easy. This feature is illustrated in Supplementary Figure \ref{supp_code_gen}.

        \begin{figure}[h!]
            \centering
            \includegraphics[scale=0.60]{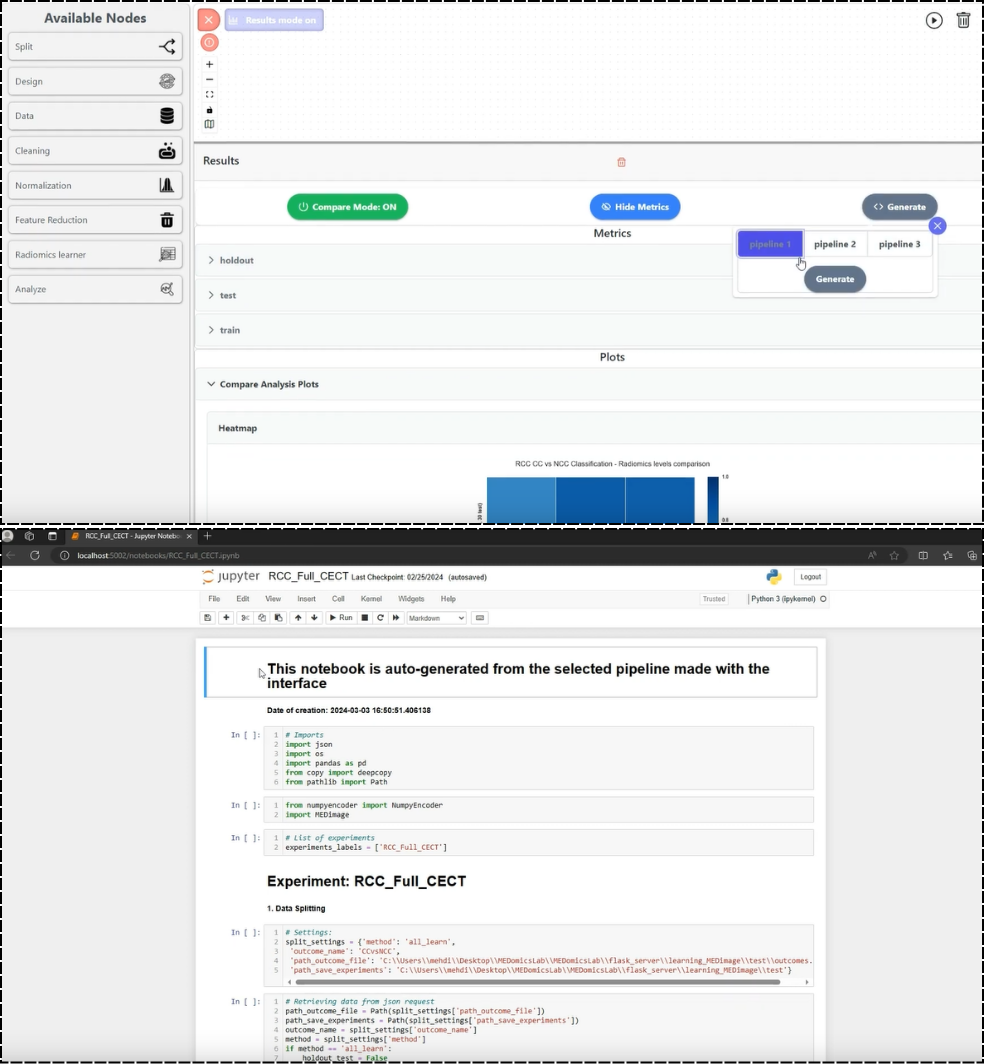}
            \caption[Code Generation for Selected Machine Learning Experiments]{Code Generation for Selected Machine Learning Experiments. Users can select and generate code for a chosen pipeline (top image). After clicking generate, a Python notebook is automatically created and opened (bottom image).}
            \label{supp_code_gen}
        \end{figure}

    \subsection*{Comparison of existing tools for radiomics research}

    We incorporated our package into the comparison conducted by Abler \textit{et al.} \cite{quantimage_v2}, making minor updates. Additionally, we introduced the feature of prediction modeling analysis to the comparison. The updated review results are provided in table \ref{tab:packages_review}.

    \begin{table}[htbp]
    \footnotesize
    \centering
    \begin{tabular}{|lllllllll|}
    \hline
    \multicolumn{1}{|l|}{\multirow{2}{*}{Name}} & \multicolumn{1}{l|}{\multirow{2}{*}{Type}} & \multicolumn{1}{l|}{\multirow{2}{*}{DICOM}} & \multicolumn{3}{c|}{Radiomics features} & \multicolumn{3}{l|}{Predictive modeling} \\ \cline{4-9} 
    \multicolumn{1}{|l|}{} & \multicolumn{1}{l|}{} & \multicolumn{1}{l|}{} & \multicolumn{1}{l|}{Extraction} & \multicolumn{1}{l|}{Visualization} & \multicolumn{1}{l|}{Selection} & \multicolumn{1}{m{0.5cm}|}{\rotatebox{-90}{Training}} & \multicolumn{1}{m{0.5cm}|}{\rotatebox{-90}{Evaluation}} & \multicolumn{1}{m{0.5cm}|}{\rotatebox{-90}{Analysis}} \\ \hline
    \multicolumn{1}{|l|}{\begin{tabular}[c]{@{}l@{}}PyRadiomics**\\ moddicom$^+$\\ RADIOMICS\\ PORTS\\ ROdiomiX*\\ SERA**$^+$\\ QIFE**\\ MIRP**$^+$\\ RaCaT\\ Precision-medicine-toolbox\\ LIFEx**$^+$\\ S-IBEx**$^+$\\ CERR**$^+$\\ MRP\\ MITK Phenotyping**\\ SlicerRadiomics\\ CGITA\\ QuantImage (v1)\\ ePAD\\ MaZda/b11\\ CaPTk**$^+$\\ AutoRadiomics**\\ QuantImage v2**\\ MEDimage**$^+$\end{tabular}} & \multicolumn{1}{l|}{\begin{tabular}[c]{@{}l@{}}Library\\ Library\\ Library\\ Library\\ cmd-exec\\ Library/\\ Library/\\ Library\\ cmd-exec\\ Library\\ GUI\\ GUI\\ GUI(plugin)\\ GUI(plugin)\\ GUI(plugin)\\ GUI(plugin)\\ GUI(plugin)\\ GUI(web)\\ GUI(web)\\ GUI\\ GUI\\ GUI(web)\\ GUI(web)\\ Library/GUI\end{tabular}} & \multicolumn{1}{l|}{\begin{tabular}[c]{@{}l@{}}-\\ x\\ x\\ -\\ x\\ -\\ x\\ x\\ x\\ x\\ x\\ x\\ x\\ x\\ x\\ x\\ x\\ x\\ x\\ -\\ x\\ x\\ x\\ x\end{tabular}} & \multicolumn{1}{l|}{\begin{tabular}[c]{@{}l@{}}x\\ x\\ x\\ x\\ x\\ x\\ x\\ x\\ x\\ PyRadiomics\\ x\\ x\\ x\\ x\\ x\\ PyRadiomics\\ x\\ x\\ QIFE\\ x\\ x\\ x\\ x\\ x\end{tabular}} & \multicolumn{1}{l|}{\begin{tabular}[c]{@{}l@{}}\\  \\  \\ \\ \\ \\ \\ \\ \\ Cohort\\ Feature map\\ -\\ Feature map\\ -\\ -\\ -\\ -\\ -\\ Feature map\\ Feature map\\ Feature map\\ Feature map\\ Cohort\\ Cohort\end{tabular}} & \multicolumn{1}{l|}{\begin{tabular}[c]{@{}l@{}} \\  \\  \\ \\ \\ \\  \\  \\ \\ \\ \\  \\  \\ \\ \\ \\  \\  \\ \\ Automated\\ -\\ Automated\\ Interactive\\ Automated\end{tabular}} & \multicolumn{1}{l|}{\begin{tabular}[c]{@{}l@{}} \\  \\  \\ \\ \\ \\ \\ \\ \\ x \\ \\ \\ \\ \\ \\ \\ \\ \\ \\ x\\ x\\ x\\ x\\ x\end{tabular}} & \multicolumn{1}{l|}{\begin{tabular}[c]{@{}l@{}} \\  \\  \\ \\ \\ \\ \\ \\ \\ x\\ \\ \\ \\ \\ \\ \\ \\ \\ \\ -\\ x\\ x\\ x\\ x\end{tabular}} & \begin{tabular}[c]{@{}l@{}} \\  \\  \\ \\ \\ \\ \\ \\ \\  -\\ \\ \\ \\ \\ \\ \\ \\ \\ \\ -\\ -\\ -\\ -\\ x\end{tabular} \\ \hline
    \multicolumn{9}{|l|}{\begin{tabular}[c]{@{}l@{}}Cross (x) or dash (-) indicate the presence or absence, respectively, of a specific characteristic. \\ Features not typically applicable to a specific tool category are left empty. \\ *Participation in the IBSI-1 benchmark \cite{ibsi_1}.\\$^+$Participation in the IBSI-2 benchmark \cite{ibsi_2}. \\ **Self-reported adherence to IBSI-1 recommendations \cite{ibsi_1}. \\ cmd-exec: Command line executable.\\ GUI: graphical user interface.\\ IBSI: Image Biomarker Standardisation Initiative. \end{tabular}} \\ \hline
    \end{tabular}
    \caption{Review summary of open-source radiomics tools.}
    \label{tab:packages_review}
    \end{table}

\section*{Supplementary Note 3: Dataset}
\label{sec:supp_datasets}
    Five distinct datasets, each associated with a different medical context have been used in this study. All the cohorts characteristics are summarized in the tables below:

    \begin{itemize}
        \item Non-Small Cell Lung Cancer (NSCLC) dataset for histology classification: The data utilized for histological classification of Non-Small Cell Lung Cancer (NSCLC) was sourced from Primakov \textit{et al.} study \cite{nsclc_study} and consisted of three institutions. Only patient cohort treated at the MAASTRO clinic, the netherlands, between 2005 and 2010 and at the Stanford University Medical Center between 2008 and 2012 are publicly available and accessible via The Cancer Imaging Archive (only patients with non-contrast-enhanced CT scans were retained) at \url{https://doi.org/10.7937/K9/TCIA.2015.PF0M9REI} and \url{https://doi.org/10.7937/K9/TCIA.2017.7hs46erv}. Patients from the University of California San Francisco (UCSF) are not publicly shared. The scans were accompanied by segmentation maps outlining tumor regions, hence no further preprocessing was performed. Patient information is listed in table \ref{tab:lungs_cohort}.

        \begin{table}[htbp]
        \footnotesize
        \centering
        \begin{tabular}{|llc|}
        \hline
        \multicolumn{1}{|l|}{Characteristics} & \multicolumn{1}{l|}{Type} & \multicolumn{1}{l|}{No of patients (\%) / value} \\ \hline
        \multicolumn{1}{|l|}{Gender}      & \multicolumn{1}{l|}{\begin{tabular}[c]{@{}l@{}} Male – MAASTRO \\ Male – Stanford \\ Male – UCSF \\ Male – All \\ Female – MAASTRO \\ Female – Stanford \\ Female – UCSF \\ Female – All \\ Missing (UCSF) \end{tabular}} & \begin{tabular}[c]{@{}c@{}} 136 (66 \%) \\ 102 (75 \%) \\ 75 (46 \%) \\ 313 (62 \%) \\ 71 (34 \%) \\ 34 (25 \%) \\ 87 (54 \%) \\ 192 (38 \%) \\ 1 \end{tabular} \\ \hline
        \multicolumn{1}{|l|}{Age}      & \multicolumn{1}{l|}{\begin{tabular}[c]{@{}l@{}} MAASTRO \\ UCSF \\ Stanford \\ All \end{tabular}} & \begin{tabular}[c]{@{}c@{}} 69 ± 10 [69; 43-92] years \\ 69 ± 9 [69; 43-87] years \\ 70 ± 9 [72; 46-92] years \\ 69 ± 9 [70; 43-92] years \end{tabular} \\ \hline
        \multicolumn{1}{|l|}{Histology} & \multicolumn{1}{l|}{\begin{tabular}[c]{@{}l@{}} Adenocarcinoma – MAASTRO \\ Adenocarcinoma – Stanford \\ Adenocarcinoma – UCSF \\ Adenocarcinoma – All \\ Other – MAASTRO \\ Other – Stanford \\ Other – UCSF \\ Other - All \end{tabular}} & \begin{tabular}[c]{@{}c@{}} 34 (16 \%) \\ 106 (78 \%) \\ 100 (61 \%) \\ 240 (47 \%) \\ 173 (84 \%) \\ 30 (22 \%) \\ 63 (39 \%) \\ 266 (53 \%) \end{tabular} \\ \hline
        \multicolumn{1}{|l|}{Treatment - MAASTRO}      & \multicolumn{1}{l|}{\begin{tabular}[c]{@{}l@{}} Radiotherapy only \\ Chemo-radiotherapy \end{tabular}} & \begin{tabular}[c]{@{}c@{}} 46.5 \% \\ 53.5 \% \end{tabular} \\ \hline
        \multicolumn{1}{|l|}{Treatment - Stanford}      & \multicolumn{1}{l|}{\begin{tabular}[c]{@{}l@{}} Surgery \\ Radiotherapy \\ Chemotherapy \\ Adjuvant therapy \end{tabular}} & \begin{tabular}[c]{@{}c@{}} 100 \% \\ 36 \% \\ 12 \% \\ 36 \%  \end{tabular} \\ \hline
        \multicolumn{1}{|l|}{Treatment - UCSF}      & \multicolumn{1}{l|}{\begin{tabular}[c]{@{}l@{}} Surgery \\ Radiotherapy \\ Chemotherapy \\ Immunotherapy  \end{tabular}} & \begin{tabular}[c]{@{}c@{}} 63 \% \\ 57 \% \\ 52 \% \\ 5 \% \end{tabular} \\ \hline
        
        \multicolumn{3}{|l|}{\begin{tabular}[c]{@{}l@{}} Note: mean ± std [median; min-max]. \\ UCSF: University California San Francisco.
        \end{tabular}} \\ \hline
        \end{tabular}
        \caption{Patient information – Lung cancer cohort.}
        \label{tab:lungs_cohort}
        \end{table}

        \item Low grade glioma (LGG): The dataset employed for the prediction of IDH1 mutation and comprises two cohorts. The first cohort (n = 227) was treated at the Department of Neurosurgery of Huashan hospital between 2010 and 2016, and created from the combined studies of Yu \textit{et al.} \cite{lgg_dataset_14} and Li \textit{et al.} \cite{lgg_dataset_15}. Outcome information and imaging data is available from the authors of these studies upon request. The second cohort (n=107) was part of The Cancer Genome Atlas Low Grade Glioma (TCGA-LGG) \cite{lgg_tcga_13}. Outcome information as well as clinical, imaging and genomics data is available from TCIA at \url{http://doi.org/10.7937/K9/TCIA.2016.L4LTD3TK}. Patient information is listed in table \ref{tab:lgg_cohort}.

        \begin{table}[htbp]
        \footnotesize
        \centering
        \begin{tabular}{|llc|}
        \hline
        \multicolumn{1}{|l|}{Characteristics} & \multicolumn{1}{l|}{Type} & \multicolumn{1}{l|}{No of patients (\%) / value} \\ \hline
        \multicolumn{1}{|l|}{Gender}      & \multicolumn{1}{l|}{\begin{tabular}[c]{@{}l@{}} Male – TCGA \\ Female – TCGA  \end{tabular}} & \begin{tabular}[c]{@{}c@{}} 50 (47 \%) \\ 57 (53 \%) \end{tabular} \\ \hline
        \multicolumn{1}{|l|}{Age}      & \multicolumn{1}{l|}{\begin{tabular}[c]{@{}l@{}} TCGA \end{tabular}} & \begin{tabular}[c]{@{}c@{}} 46 ± 14 [47; 20-75] years \end{tabular} \\ \hline
        \multicolumn{1}{|l|}{Laterality} & \multicolumn{1}{l|}{\begin{tabular}[c]{@{}l@{}} Left – TCGA \\ Midline – TCGA \\ Right – TCGA \end{tabular}} & \begin{tabular}[c]{@{}c@{}} 48 (45 \%) \\ 3 (3 \%) \\56 (52 \%) \end{tabular} \\ \hline
        \multicolumn{1}{|l|}{Tumour location}      & \multicolumn{1}{l|}{\begin{tabular}[c]{@{}l@{}} Cerebellum – TCGA \\ Frontal – TCGA \\ Parietal – TCGA \\ Temporal – TCGA \\ Not specified – TCGA \\ Missing \end{tabular}} & \begin{tabular}[c]{@{}c@{}} 1 (1 \%) \\ 58 (58 \%) \\ 12 (12 \%) \\ 28 (28 \%) \\1 (1 \%) \\ - \end{tabular} \\ \hline
        \multicolumn{1}{|l|}{IDH1 - mutation}      & \multicolumn{1}{l|}{\begin{tabular}[c]{@{}l@{}} Yes – Huashan \\ Yes – TCGA \\ Yes – All \\ No – Huashan \\ No – TCGA \\ No – All \\ Missing (TCGA) \end{tabular}} & \begin{tabular}[c]{@{}c@{}} 164 (72 \%) \\ 76 (74 \%) \\ 240 (73 \%) \\ 63 (28 \%) \\ 27 (26 \%) \\ 90 (27 \%) \\ 4  \end{tabular} \\ \hline
        \multicolumn{1}{|l|}{Treatment - TCGA}      & \multicolumn{2}{l|}{See \url{http://doi.org/10.7937/K9/TCIA.2016.L4LTD3TK}} \\ \hline
        \multicolumn{1}{|l|}{Treatment - Huashan}      & \multicolumn{2}{l|}{Not available} \\ \hline
        
        \multicolumn{3}{|l|}{\begin{tabular}[c]{@{}l@{}} Note: mean ± std [median; min-max]. \\ UCSF: University California San Francisco.
        \end{tabular}} \\ \hline
        \end{tabular}
        \caption{Patient information – Low-grade-glioma cohort.}
        \label{tab:lgg_cohort}
        \end{table}

        \item Meningioma (MRI): The dataset employed for pathological grade classification comprises two cohorts. The first cohort (n=257) includes patients treated at the Radiation Oncology Department of UCSF between 2001 and 2013, sourced from studies by Wu \textit{et al.} \cite{wu_presenting_2018}, Vasudevan \textit{et al.} \cite{vasudevan_comprehensive_2018}, Gennatas \textit{et al.} \cite{gennatas_preoperative_2018}, and Morin \textit{et al.} \cite{meningioma_dataset_16}. The second cohort (n=87) comprises patients treated at Princess Margaret Hospital (PMH) in Toronto between 2010 and 2017, as detailed in the study by Morin \textit{et al.} \cite{meningioma_dataset_16}. The dataset is not publicly shared by the hosting institutions. Patient information is listed in table \ref{tab:meningioma_cohort}.

        \begin{table}[htbp]
        \footnotesize
        \centering
        \begin{tabular}{|llc|}
        \hline
        \multicolumn{1}{|l|}{Characteristics} & \multicolumn{1}{l|}{Type} & \multicolumn{1}{l|}{No of patients (\%) / value} \\ \hline
        \multicolumn{1}{|l|}{Gender} & \multicolumn{1}{l|}{\begin{tabular}[c]{@{}l@{}}Male – UCSF \\ Male – PMH \\ Male – All \\ Female – UCSF \\ Female – PMH \\ Female – All \end{tabular}} & \begin{tabular}[c]{@{}c@{}} 96 (37 \%) \\ 31 (36 \%) \\ 127 (37 \%) \\161 (63 \%) \\ 56 (64 \%) \\ 217 (63 \%)  \end{tabular} \\ \hline
        \multicolumn{1}{|l|}{Age}      & \multicolumn{1}{l|}{\begin{tabular}[c]{@{}l@{}} UCSF \\ PMH \\ All \end{tabular}} & \begin{tabular}[c]{@{}c@{}} 58 ± 13 [58; 14-89] years \\ 58 ± 15 [59; 19-88] years \\ 58 ± 14 [58; 14-89] years \end{tabular} \\ \hline
        \multicolumn{1}{|l|}{Pathology – Grade}      & \multicolumn{1}{l|}{\begin{tabular}[c]{@{}l@{}} Grade 1 – UCSF \\ Grade 1 – PHM \\ Grade 1 – All \\ Grade 2 – UCSF \\ Grade 2 – PMH \\ Grade 2 – All \\ Grade 3 – UCSF \\ Grade 3 – PMH \\ Grade 3 – All \end{tabular}} & \begin{tabular}[c]{@{}c@{}} 128 (50 \%) \\ 69 (79 \%) \\ 197 (57 \%) \\ 104 (40 \%) \\ 17 (20 \%) \\ 121 (35 \%) \\ 25 (10 \%) \\ 1 (1 \%) \\ 26 (8 \%) \end{tabular} \\ \hline
        \multicolumn{1}{|l|}{Tumour location}      & \multicolumn{1}{l|}{\begin{tabular}[c]{@{}l@{}} Cerebellum – TCGA \\ Frontal – TCGA \\ Parietal – TCGA \\ Temporal – TCGA \\ Not specified – TCGA \\ Missing \end{tabular}} & \begin{tabular}[c]{@{}c@{}} 1 (1 \%) \\ 58 (58 \%) \\ 12 (12 \%) \\ 28 (28 \%) \\ 1 (1 \%) \\ - \end{tabular} \\ \hline
        \multicolumn{1}{|l|}{IDH1 mutation}      & \multicolumn{1}{l|}{\begin{tabular}[c]{@{}l@{}} Yes – Huashan \\ Yes – TCGA \\ Yes – All \\ No – Huashan \\ No – TCGA \\ No – All \\ Missing (TCGA) \end{tabular}} & \begin{tabular}[c]{@{}c@{}} 164 (72 \%) \\ 76 (74 \%) \\ 240 (73 \%) \\ 63 (28 \%) \\ 27 (26 \%) \\ 90 (27 \%) \\ 4 \end{tabular} \\ \hline
        \multicolumn{1}{|l|}{Treatment - UCSF}      & \multicolumn{1}{l|}{\begin{tabular}[c]{@{}l@{}} Extent of resection: \\ - Gross total resection \\ - Subtotal resection \\  Adjuvant radiotherapy \end{tabular}} & \begin{tabular}[c]{@{}c@{}} \\ 56 \% \\ 44 \% \\ 24 \%\end{tabular} \\ \hline
        \multicolumn{1}{|l|}{Treatment - PMH}      & \multicolumn{1}{l|}{\begin{tabular}[c]{@{}l@{}} Extent of resection: \\ - Gross total resection \\ - Subtotal resection \\ Unknown \\ Adjuvant radiotherapy \end{tabular}} & \begin{tabular}[c]{@{}c@{}}  \\ 56 \% \\ 12 \% \\ 32 \% \\ 5 \% \end{tabular} \\ \hline
        \multicolumn{3}{|l|}{\begin{tabular}[c]{@{}l@{}} Note: mean ± std [median; min-max]. \\ For binary prediction of pathological grade:\\ - Grade 1 is considered as “Low” (0) \\ - Grade 2 and 3 are considered as “High” (1). \\ UCSF: University California San Francisco. \\ PMH: Princess Margaret Hospital.
        \end{tabular}} \\ \hline
        \end{tabular}
        \caption{Patient information – Meningioma cohort.}
        \label{tab:meningioma_cohort}
        \end{table}

        \item Renal Cell Carcinomas (RCC) dataset: Part of the dataset utilized for classifying Renal Lesions was obtained from the work of Ianto Lin Xi \textit{et al.} \cite{rcc_dataset_17}, while the rest was generated by the TCGA Research Network (\url{https://doi.org/10.7937/K9/TCIA.2016.V6PBVTDR}), comprising a collection of 1197 MR images, including two enhanced sequences: T1-contrast (T1C, n=598) and T2-weighted (T2WI, n=599). Manual segmentation of the images was performed by three radiologists to delineate regions of interest. Subsequently, only T2WI images were retained for analysis, as the initial investigation revealed similar results to those obtained from the NSCLC and CECT-based RCC datasets. Patient information is accessible in the dataset sources.

        \item Renal Cell Carcinomas (RCC) dataset: This dataset comprises 326 patients diagnosed with renal cell carcinoma (RCC) and treated at Centre hospitalier universitaire de Sherbrooke (CHUS), QC, Canada. Manual segmentation was performed by three residents under the supervision of a urologist. The images were acquired using Contrast Enhanced Computed Tomography (CECT) and were used to classify clear versus non-clear Cells. This dataset is not publicly shared by the institution. Patient information is listed in table \ref{tab:rcc_cect_cohort}.

        \begin{table}[htbp]
        \centering
        \begin{tabular}{|llc|}
        \hline
        \multicolumn{1}{|l|}{Characteristics} & \multicolumn{1}{l|}{Type} & \multicolumn{1}{l|}{No of patients (\%) / value} \\ \hline
        \multicolumn{1}{|l|}{Gender}      & \multicolumn{1}{l|}{\begin{tabular}[c]{@{}l@{}} Male – CHUS \\ Female – CHUS \end{tabular}} & \begin{tabular}[c]{@{}c@{}} 217 (67 \%) \\ 109 (33 \%) \end{tabular} \\ \hline
        \multicolumn{1}{|l|}{Age}      & \multicolumn{1}{l|}{\begin{tabular}[c]{@{}l@{}} CHUS \end{tabular}} & \begin{tabular}[c]{@{}c@{}} 63.3 ± 10.53 [65; 33-91] years \end{tabular} \\ \hline
        \multicolumn{1}{|l|}{Subtype} & \multicolumn{1}{l|}{\begin{tabular}[c]{@{}l@{}} Clear cell \\ Non-Clear cell \end{tabular}} & \begin{tabular}[c]{@{}c@{}} 79 (24 \%) \\ 247 (76 \%) \end{tabular} \\ \hline
        \multicolumn{1}{|l|}{Lesion side}      & \multicolumn{1}{l|}{\begin{tabular}[c]{@{}l@{}} Left \\ Right \end{tabular}} & \begin{tabular}[c]{@{}c@{}} 170 (52 \%) \\ 156 (48 \%) \end{tabular} \\ \hline
        \multicolumn{1}{|l|}{Imaging size}      & \multicolumn{1}{l|}{\begin{tabular}[c]{@{}l@{}} CHUS \end{tabular}} & \begin{tabular}[c]{@{}c@{}} 5.14 ± 2.99 [4.5; 1-15] cm  \end{tabular} \\ \hline
        \multicolumn{1}{|l|}{Family History}      & \multicolumn{1}{l|}{\begin{tabular}[c]{@{}l@{}} Yes \\ No \end{tabular}} & \begin{tabular}[c]{@{}c@{}} 18 (5\%) \\ 308 (95\%) \end{tabular} \\ \hline
        
        \multicolumn{3}{|l|}{\begin{tabular}[c]{@{}l@{}} Note: mean ± std [median; min-max]. \\ cm: centimeter \\ CHUS: Centre hospitalier universitaire de Sherbrooke.
        \end{tabular}} \\ \hline
        \end{tabular}
        \caption{Patient information – CECT-based Renal cell carcinoma cohort.}
        \label{tab:rcc_cect_cohort}
        \end{table}

    \end{itemize}

\section*{Supplementary Note 4: Visualization of the impact of features with the highest importance}
\label{sec:supp_visualization}

    Additional figures were included here, each showcasing two representative images corresponding to the two classes of the binary problem studied. These images were automatically chosen from patients with the highest difference in the feature with the highest importance during the model's training process.

    \begin{figure}[h!]
        \centering
        \includegraphics[scale=0.4]{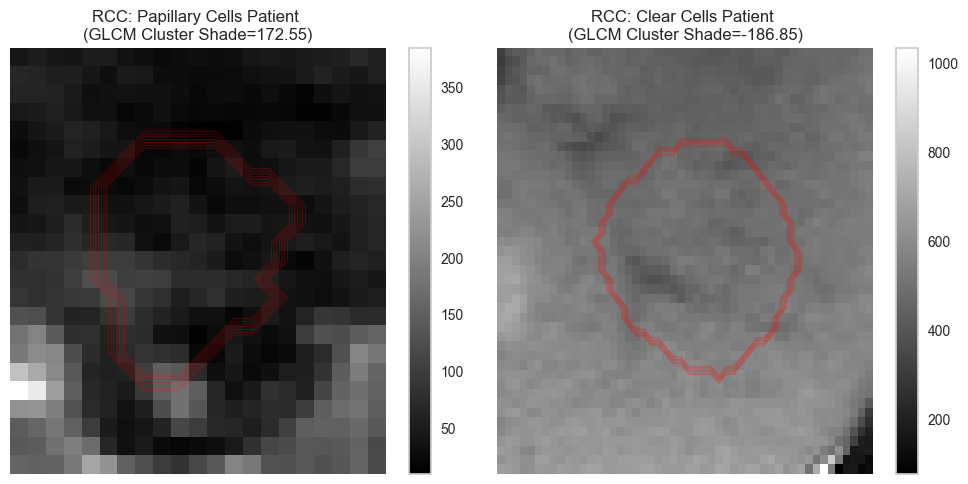}
        \caption[Axial MRI comparison of clear versus papillary cells patients with the highest disparity in the texture feature cluster shade.]{Axial MRI comparison of clear versus papillary cells patients with the highest disparity in the texture feature cluster shade.}
    \end{figure}

    \begin{figure}[h!]
        \centering
        \includegraphics[scale=0.4]{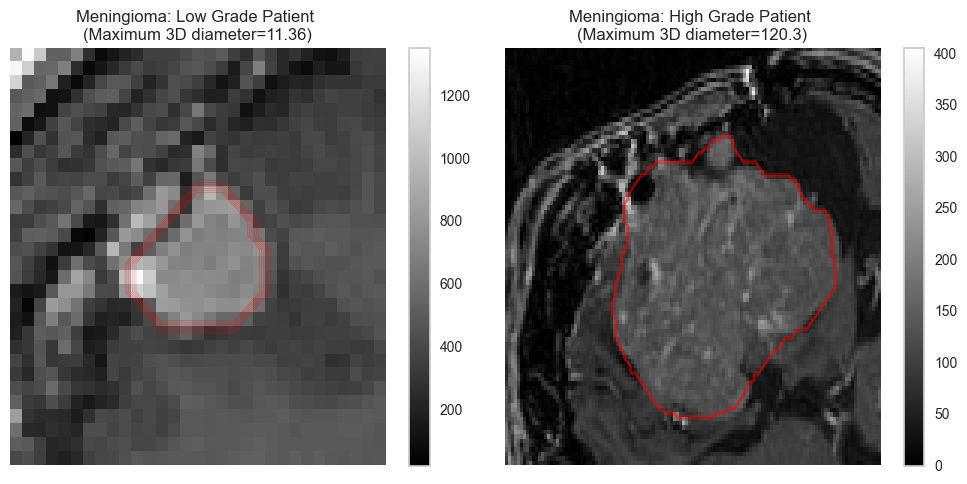}
        \caption[Axial Magnetic Resonance Imaging (MRI) comparison of low grade versus high grade patients with the highest disparity in the morphological feature maximum 3D diameter.]{Axial Magnetic Resonance Imaging (MRI) comparison of low grade versus high grade patients with the highest disparity in the morphological feature maximum 3D diameter.}
    \end{figure}

    \begin{figure}[h!]
        \centering
        \includegraphics[scale=0.4]{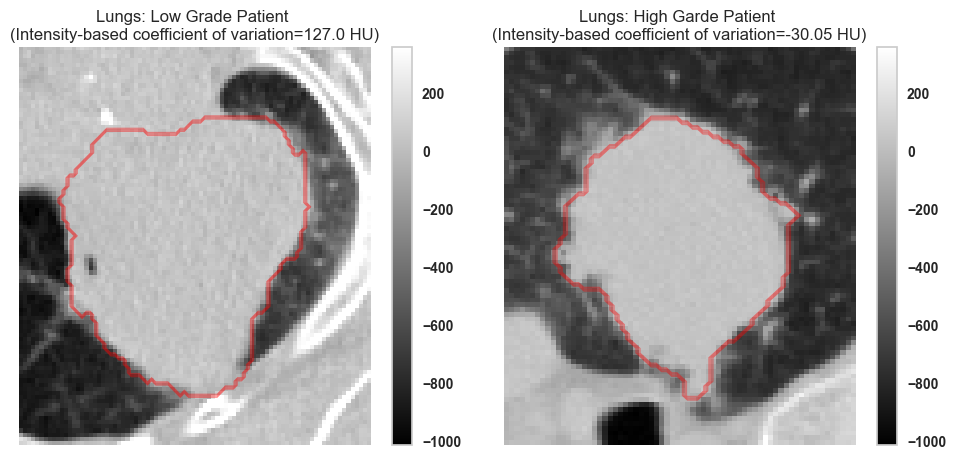}
        \caption[Axial Computed Tomography (CT) comparison of low grade versus high grade patients with the highest disparity in the intensity coefficient of variation.]{Axial Computed Tomography (CT) comparison of low grade versus high grade patients with the highest disparity in the intensity coefficient of variation.}
    \end{figure}

    \begin{figure}[ht!]
        \centering
        \includegraphics[scale=0.4]{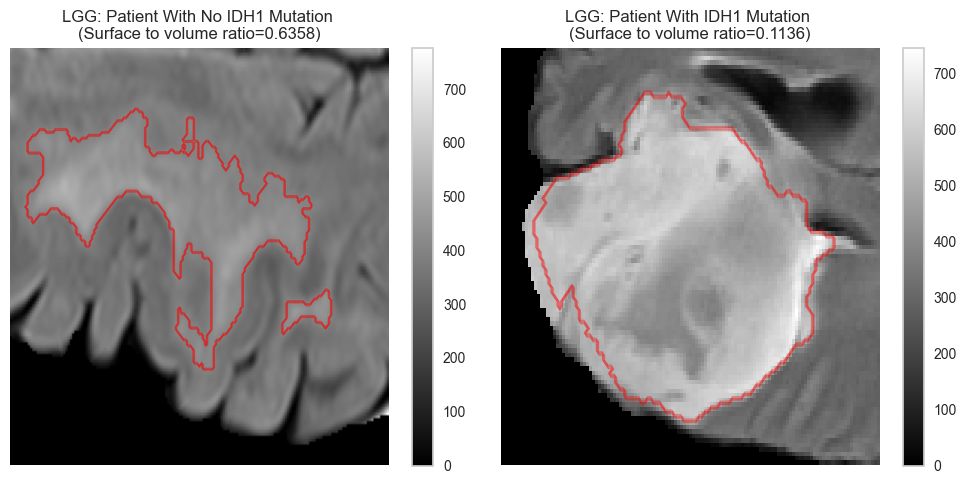}
        \caption[Axial MRI comparison of patients with and without the IDH-1 mutation with the highest disparity in the morphological ratio surface to volume.]{Axial MRI comparison of patients with and without the IDH-1 mutation with the highest disparity in the morphological ratio surface to volume.}
    \end{figure}

\section*{Supplementary Note 5: Highlighting the potential of textural filters in Lungs cohort}
\label{sec:supp_tf_analysis}

    To our best knowledge, our work is the first to leverage the textural filters for extracting radiomics features. In the histological subtypes classification of lung cancer based on CT, features derived from textural filters matched the AUC of the selected optimal level (0.76; 95\% CI: 0.72, 0.80) which was the intensity features based model, the sensitivity (0.73), but with a specificity 3\% higher, underscoring their considerable potential.
    
    To identify filters, feature families and features that contribute most to the decision-making process for a given complexity level, we used feature importance, a measure of how much each feature contributes to the predictive power of the model. In XGBoost, feature importance is computed based on how often and how significantly a feature is used to split data across all trees in the model. We used this measure and created the feature importance tree (see Supplementary Figure \ref{fig:tf_tree}), a plot that breaks down the complexity level in a hierarchical fashion. The tree branches from top to bottom, starting with the filter type used (if applicable), followed by the filter name (if applicable), then the feature families for texture features, and finally, individual features. Branch thickness (green lines) indicates the relative importance of different features or feature groups. Solid lines show selected features or paths, while black dotted lines represent features included in the set but not selected for the final model. The orange line traces the path from the root to the leaf node with the highest accumulated feature importance in the model's decision-making process. According to the plot, the Cluster Shade filter and the High Dependence Low Gray Level Emphasis (HDLGE) feature had the highest importance.

    \begin{figure}[h!]
        \centering
        \includegraphics[scale=0.38]{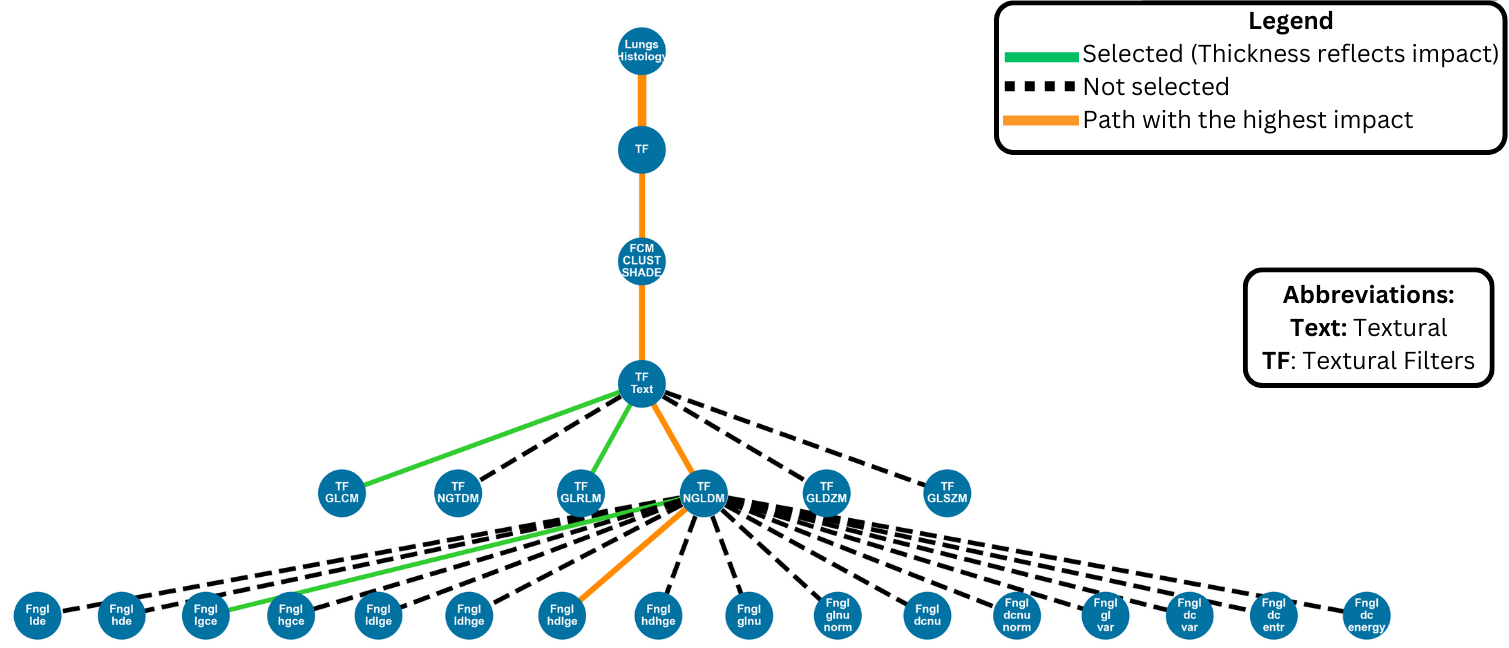}
        \caption[Feature importance tree depicting the textural filters complexity level]{Feature importance tree depicting the textural filters complexity level and showing the filter and features with the highest importance: Cluster shade-based filter and high dependence low gray level emphasis (hdlge) feature.}
        \label{fig:tf_tree}
    \end{figure}

\section*{Supplementary media}
\label{sec:supp_media}

\begin{itemize}
    \item A MEDimage promotional video to highlight how this platform allows the graphical customization of radiomics studies through node movement and linking: \url{https://youtu.be/h38vEpkHSpc?feature=shared}
    
    \item The MEDimage package documentation: \url{https://medimage.readthedocs.io}
\end{itemize}

\begin{itemize}
    \item The MEDimage app documentation: \url{https://medomics-udes.gitbook.io/medimage-app-docs}
\end{itemize}

\bibliographystyle{unsrtdin}  

\end{document}